\documentclass[twocolumn]{aastex631}

\usepackage{amsmath,amssymb}
\usepackage{verbatim}
\usepackage{color}
\usepackage{hyperref}
\usepackage{booktabs}
\usepackage{float}
\usepackage{needspace}
\usepackage{graphicx}
\usepackage{enumitem}
\usepackage{halloweenmath}
\usepackage{natbib}

\definecolor{linkcolor}{rgb}{0,0,0.25}
\hypersetup{
  colorlinks=true,        
  linkcolor=linkcolor,    
  citecolor=linkcolor,    
  filecolor=linkcolor,    
  urlcolor=linkcolor      
}

\newcounter{address}
\DeclareMathAlphabet{\mathsc}{OT1}{cmr}{m}{sc}
\def\testbx{bx}%
\DeclareRobustCommand{\ion}[2]{%
  \relax\ifmmode
  \ifx\testbx\f@series
    {\mathbf{#1\,\mathsc{#2}}}\else
    {\mathrm{#1\,\mathsc{#2}}}\fi
  \else\textup{#1\,{\mdseries\textsc{#2}}}%
  \fi
}
\usepackage{xcolor}

\newcommand{\mvr}{$\bar v_{\mathrm{R}}$}

\begin{document}

\title{The Milky Way’s rowdy neighbours: The effects of the Large Magellanic Cloud and Sagittarius Dwarf on the Milky Way Disc}

\author[0000-0002-4110-8769]{Ioana A. Stelea} \email{stelea@wisc.edu}
\affiliation{Department of Astronomy, University of Wisconsin–Madison, Madison, WI, 53706}

\author [0000-0001-8917-1532]
{Jason A.~S.~Hunt}
\affiliation{School of Mathematics \& Physics, University of Surrey, Guildford, GU2 7XH, UK}

\author[0000-0001-6244-6727]{Kathryn V. Johnston}
\affiliation{Department of Astronomy, Columbia University, New York, NY 10027, USA,}

\title[MW's rowdy neighbours]
{The Milky Way’s rowdy neighbours: The effects of the Large Magellanic Cloud and Sagittarius Dwarf on the Milky Way Disc}

\begin{abstract}

  The Milky Way (MW) is a barred spiral galaxy shaped by tidal interactions with its satellites. The Large Magellanic Cloud (LMC) and the Sagittarius Dwarf galaxy (Sgr) are the dominant influences at the present day. This paper presents a suite of four $10^9$ particle N-body simulations, illustrating the response of the stellar disc of the MW to the close approach of the LMC and the merger of Sgr into the MW. The suite is intended to provide a resource for others to study the complex interactions between the MW and its satellites independently and together, in comparison to an isolated disc control simulation. The high temporal and mass resolution allows for a quantitative Fourier decomposition of the stellar kinematics, disentangling the individual influence of each satellite on the MW. In our preliminary analysis, we find that the influences from the LMC and Sgr on the disc of the MW appear distinct, additive, and separable within our tailored simulations. Notably, the corrugations induced by Sgr reproduce the large radial velocity wave seen in the data \citep{Eilers+20}. Overall, our findings emphasise the need to include both satellites when modelling the present-day state of the MW structure and kinematics.

\end{abstract}

\keywords{Galaxy: disc --- Galaxy:
kinematics and dynamics --- Galaxy: structure}

\section{Introduction}\label{intro}
The Milky Way (MW) has long been known to be a barred spiral galaxy \citep[e.g.][]{Blitz+Spergel91,Weinberg92}. The MW disc is also known to contain vertical corrugations and ripples both local to the Sun, \citep[][]{Widrow+12_verticalwaves,Williams+2013} and towards the outer disc \citep{Xu+15_corrugations,Price-Whelan+15}. Recent stellar surveys such as $Gaia$ \citep{GaiaMission} and SDSS-IV APOGEE \citep{MAPOGEE16} have let us map these structures in stellar number counts, kinematics and chemistry \citep[e.g.][]{BLHM+19,Eilers+20,Eilers+22,GaiaCollab+23_mapasym,DR3_chemcart}. $Gaia$ has also revealed new disequilibrium features in the disc such as the $z-v_z$ phase spiral \citep{Antoja+18} and the ridges and ripples in $R-v_{\phi}$ \citep{KBCCGHS18,Antoja+18}. These asymmetrical features show that our Galaxy is out of equilibrium, yet the origin of such features has many possible explanations. 

Our understanding of the MW’s structure is limited by our position within the Galaxy, and selection effects such as from the dust extinction which obscures much of the Galactic disc. Thus, we rely on theoretical models and large numerical simulations to compensate for these biases, and to help us reach a more complete understanding of the Galaxy and its structure, both in terms of internal dynamics, and the interaction between the Milky Way and its satellites. 

The LMC is now widely thought to be both relatively massive \citep[e.g. $1.4 - 1.8\times10^{11}$;][]{Besla+10,Penarubia2016,Erkal+19} and on first infall to the MW \citep{Besla+07}, slightly past its first pericentre \citep[although see][for alternative interpretations]{donghia2016ARA&A..54..363D,Vasiliev23}. This is in sharp contrast to the early models of the LMC which favored a lighter satellite which had experienced several orbits around the MW \citep[e.g. $\sim0.5-2\times10^{10}\ M_{\odot}$][]{Heller+94,Gardiner+96,vanderMarel+02}. This revision of our understanding of the mass and orbit of the LMC has motivated ongoing re-examination of the LMC's influence on the MW. 

The interaction will distort the dark matter halo of both the LMC and the MW \citep{Lilleengen+23}, leave a dark matter wake in the MW halo \citep{Garavito-Camargo+19} which can be informative on the nature of dark matter \citep{Foote+23}. The LMC also induces coherent motion in smaller MW satellites including the clustering of their orbital poles \citep{Garavito-Camargo+21,Garavito-Camargo+23}, and it induces a reflex motion in the Milky Way \citep[e.g.][]{gomez2015ApJ...802..128G} which has been observed in the velocity distribution in the stellar halo \citep{Petersen+Penarrubia21}. The LMC will also have significant consequences for the structure of the galactic disc, both directly, and as a result of torque from the distorted dark halo \citep[e.g.][]{Gomez+16}. For example, \cite{Weinberg+Blitz06} showed that the infall of the LMC can reproduce the warp in the H\texttt{I} \citep{Levine+06}, and $N$-body simulations show that it can produce the warp in the stellar disc \citep[e.g.][]{Laporte+18a_LMC,Laporte+18b_Sgr+LMC}, although the warp could also be caused by a tilted triaxial halo \citep{Han+23_tiltwarp,Han+24}.

In contrast, Sgr has experienced several pericentric passages and disc crossings since an initial infall estimated around 6 Gyr ago \citep{RuizLara+20,Das+23}, and the debris from its ongoing disruption forms streams that extend across a large portion of the sky, wrapping around our Galaxy multiple times. While the present day remnant is comparatively light \citep[$\sim4\times10^8$ M$_{\odot}$;][]{Frinchaboy+12,Vasiliev+20} there is evidence that the progenitor was considerably more massive \citep[e.g.][]{Gibbons+17}, and these repeated interactions would have a significant effect on the Milky Way's disc and halo. For example, \cite{Laporte+18b_Sgr+LMC,Laporte+19_Gaia} show qualitatively that a Sgr-like merger could be responsible for both the $z-v_z$ phase spirals and the ridges in $R-v_{\phi}$ in the Solar neighborhood, the vertical corrugations \citep{gomez2013MNRAS.429..159G} and flaring of the disc, and the numerous `feathers' in the outer disc \citep{Laporte+22_feathers}, which consist of disc stars perturbed by interaction with Sgr \citep[see also][]{Khanna+19,Hunt+21}. However, \cite{BBH21} found that a range of models for Sgr alone is unable to reproduce the amplitude and wavelength of the vertical perturbation, arguing for a more complex solution.

Such simulations are invaluable in bridging the gap between theoretical understanding and observational evidence, where self-consistent merger simulations have the potential to capture the physics of a three way interaction between massive galaxies. For example, the three system merger simulations of \cite{Laporte+18b_Sgr+LMC} and subsequent analysis \citep[][]{GrionFilho+20,Poggio+21} show that the nature of the perturbation from each satellite is relatively distinct, owing to the difference in their orbital histories. The repeated passages of Sgr produces ripples and corrugations, and the LMC produces a warp, also boosting the amplitude of the perturbations associated with Sgr.

Thus, a `complete' model of the Milky Way should also contain at least the two largest of the Milky Way's satellites, the Large Magellanic Cloud (LMC) and the Sagittarius dwarf galaxy (Sgr) which are both massive enough and close enough to tidally perturb the Milky Way disc and halo, and create some, or all, of the large scale disequilibrium features observed in the $Gaia$ data. In addition, the presence of the LMC will change the orbit of Sgr, and must be included to best reproduce the orbit of Sgr and its tidal stream \citep{gomez2015ApJ...802..128G,Vasiliev+21}. 

While there are many studies of individual Sgr and LMC mergers, the difficulty in setting up the initial conditions for a three-system merger that studies the combined effect of both satellites, means there are few recently available in the literature \citep[see e.g.][]{Laporte+18b_Sgr+LMC,Vasiliev+21}. The study of \cite{Vasiliev+21} focuses on, and achieves an excellent recovery of the Sgr stream when the influence of the LMC is included, yet their disc model is low resolution by today's standards ($10^6$ particles) and dynamically hot, used only for its contribution to the potential. In contrast, the fully self-consistent model of \cite{Laporte+18b_Sgr+LMC} nicely demonstrates the overall nature and differences in the disc response, although secular evolution of the disc also results in significant structure and kinematics not directly related to the interaction with the satellites. 

Thus, we present here a set of high-resolution models designed to closely match the `present day' state of the system, and isolate the influences of the satellites on an unperturbed disc, without confusion from secular effects. We build upon previous works to make a high-resolution model of the last 3 Gyr of the MW - Sgr - LMC interaction, with a `warm' disc. While mixing times may be faster with a higher velocity dispersion, the signal should also remain clearly visible for longer as the disc is less susceptible to secular structure formation.

We use the GPU accelerated $N$-body code \texttt{Bonsai} \citep{Bonsai,Bonsai-242bil} to run tailored high-resolution models of the LMC and Sgr mergers, individually and combined. We choose to make use of the orbital solutions of \cite{Vasiliev+21} owing to their excellent fit to the Sgr stream. Due to advances in computational capabilities, our suite of simulations introduces a factor of $\sim$40 higher resolution than previous models, which enables both higher force resolution and studies of finer phase space structure for comparison to data from $Gaia$ and ground-based surveys. The combination of a tailored setup and the increase in spatial and temporal resolution enables studies of dynamical features on small scales, and allows a quantitative comparison of satellite signatures using Fourier analysis (this work) and using basis function expansions (Petersen et al. in prep).

In this work, we present a set of four high-resolution simulations\footnote{available on reasonable request to the authors}. In Section 2, we discuss the setup and evolution of the N-body simulations, which we make publicly available. In Section 3, we present an initial analysis of the four simulations, and discuss the effect of the satellites on the disc both separately and together. In Section 4 we relax the assumption of a warm disc and show that the signatures of the satellites remain visible in a colder disc. Finally, in Section 5 we present our conclusions and discuss possible extensions.

\begin{figure}
    \includegraphics[width= 0.5\textwidth]{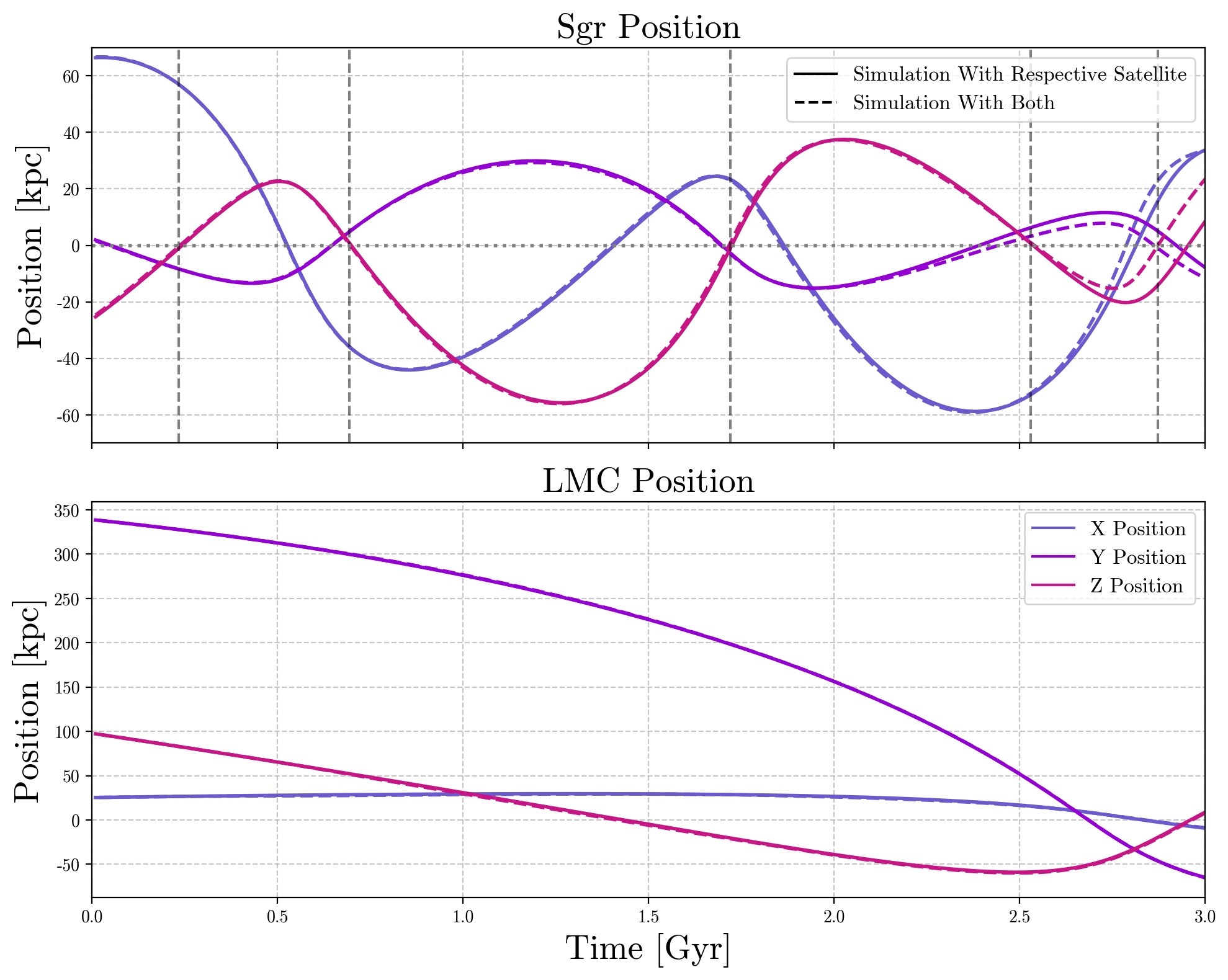}
    \caption{Orbits of Sgr (top panel) and the LMC (bottom panel), for the simulations [MW+Sgr] and [MW+LMC] containing Sgr and the LMC alone (solid lines), and the combined simulation [MW+Sgr+LMC] containing both (dashed lines).}
    \label{fig:a}
  \end{figure}

\section{The simulations}\label{sim}
In this section, we provide an overview of the setup and evolution of the four simulations: the isolated host galaxy, [MW], and the merger models, [MW+LMC], [MW+Sgr], and [MW+Sgr+LMC]. We describe how the initial conditions of the Milky Way-like disc galaxy and the two satellites were constructed and combined, and the simulation code used to evolve the systems. 

\begin{table*}
\caption{Particle number, mass and fundamental parameters of the components of the host galaxy and satellite models described in Section \ref{sim}, following \citet{Vasiliev+21}.}
\begin{tabular}{@{}llllllllll@{}}
\toprule
Component & $N_{\mathrm{p}}$ & $M_{\odot}$ & $R_d$ & $h_d$ & $R_s$ & $w_0$ & $\gamma$ & $\beta$ & $\alpha$ \\ \midrule
\midrule
Milky Way Disk & $3.2\times10^8$ & $5\times10^{10}$ & 3.0 & 0.4 & - & - & - & - & - \\
Milky Way Bulge & $8\times10^7$ & $1.2\times10^{10}$ & - & - & 0.2 & - & 0 & 1.8 & 1 \\ 
Milky Way Halo & $6\times10^8$ & $7.33\times10^{11}$ & - & - & 7.0 & - & 1 & 2.5 & 2 \\ 
\midrule
Sgr stellar & $4\times10^6$ & $2\times10^{8}$ & - & -& 1.0 & 0.4 & - & - & - \\
Sgr dark & $6.4\times10^{6}$ & $3.6\times10^{9}$ & - & - & 8.0 & - & 0 & 3 & 1 \\ 
\midrule
LMC & $2\times10^6$ & $1.5\times10^{11}$ & - & - & 10.8 & - & 1 & 3 & 1 \\
\midrule
\bottomrule
\end{tabular}
\label{ncomps}
\end{table*}

\subsection{The isolated host galaxy simulation, MW}

The initial conditions for the Milky Way-like host were generated using the galactic dynamics software package \texttt{Agama}\ \cite{agama}. We adopt the  Milky Way-like host galaxy with an axisymmetric halo from \cite{Vasiliev+21} in order to preserve the orbits of the satellites from their work. The host consists of a spherical bulge and halo and an exponential disc, the parameters, particle numbers, and mass resolution are contained in Table \ref{ncomps}. Compared to \cite{Vasiliev+21}, we reduce the disc radial velocity dispersion parameter, $\sigma_{R_0}$ (dispersion at $R=0$), from 160 to 90 km s$^{-1}$. 

This is a deliberate choice which preserves the overall mass distribution of the \cite{Vasiliev+21} model, and which sets up a disc which is bar unstable but still `hot' enough so that it forms very little other structure over 3 Gyr without the influence of the satellite galaxies. This allows us to better examine the direct influence of the satellites, rather than secular structure formation. This host galaxy is evolved alone initially as the [MW] model. We use this model to calculate the rotation required for the bar angle to lie at 25 deg with respect to the Sun--Galactic centre line at the `present day' similar to the Milky Way \citep[e.g.][]{Bland-Hawthord+Gerhard16}, assuming a Solar location of $(-8.2, 0, 0)$ kpc. We then applied this rotation to the initial conditions of the disc in the models containing satellites, as they cannot be rotated post-merger.

\begin{table*}

\caption{Table of measured position and motion of the LMC and Sgr taken from $Gaia$ \citep{Gaia_glob_dwarf+18}, \citet{Kallivayalil+13} and \citet{Vasiliev+20} compared to the `present day' position and motion of the satellites in the [MW+LMC], [MW+Sgr] and  [MW+Sgr+LMC] models.}
\noindent\makebox[\linewidth]{}
\begin{tabular}{lllllll}

\toprule
Data Source         & $x$                      & $y$                   & $z$                    & $v_x$                  & $v_y$                    & $v_z$                   \\
\toprule
LMC                 &                          &                       &                        &                        &                          &                         \\
\midrule

\cite{Gaia_glob_dwarf+18}                & $-1.078^{+0.3}_{-0.3}$   & $-41.0^{+0.2}_{-0.2}$ & $-27.82^{+1.4}_{-1.4}$ & $-60.2^{+10.2}_{-9.7}$ & $-216.6^{+13.8}_{-13.5}$ & $209.4^{+18.0}_{-18.8}$ \\
\cite{Kallivayalil+13} & -0.97                      & -41.0                   & -27.8                    & $-57^{+13}_{-13}$             & $-226^{+15}_{-15}$              & $221^{+19}_{-19}$               \\
{[}MW+LMC{]}             & -1.04      & -38.7          & -27.5               & -59.4              & -208.8                & 175.0               \\
{[}MW+Sgr+LMC{]}            & -1.09      & -40.4           & -26.8                & -61.2               & -205.6               & 177.4                \\
\midrule
Offset (wrt Gaia)   & $\Delta  x$    & $\Delta  y$        & $\Delta  z$   & $\Delta  v_x$    & $\Delta  v_y$     & $\Delta  v_z$                     \\
{[}MW+LMC{]}             & -0.03     & -2.2                  & -0.3                    & -0.7                & -7.7       & 34.3              \\
{[}MW+Sgr+LMC{]}             & 0.02    & -0.5    & -0.9    & -0.7       & -10.9     & 31.9    \\
\toprule

                    &                          &                       &                        &                        &                          &                         \\
Sagittarius                 &                          &                       &                        &                        &                          &                         \\
\midrule

\cite{Gaia_glob_dwarf+18}                & $17.0 ^{+2.0 }_{-1.8}$ & $2.5^{+0.2}_{-0.2}$   & $-6.42^{+0.5}_{-0.5}$  & $229.4^{+7.2}_{-6.2}$  & $-14.7^{+19.9}_{-22.5}$   & $205.8^{+18.6}_{-17.1}$ \\
\cite{Vasiliev+20} & 17.5       & 2.5                   & -6.5                   & 237.9                  & -24.3                    & 209.0                     \\
{[}MW+Sgr{]}             & 18.4       & 2.6                & -12.3        & 206.4      & -108.4               & 151.5
\\
{[}MW+Sgr+LMC{]}              & 16.2   & 1.6                & -6.4      & 236.9      & -98.9       & 183.5               \\
\midrule
Offset (wrt Gaia)             & $\Delta  x$     & $\Delta  y$      & $\Delta  z$     & $\Delta  v_x$    & $\Delta  v_y$     & $\Delta  v_z$                     \\
{[}MW+Sgr{]}             & -1.4          & -0.1      & 5.9        & 23.0       & 93.4       & 54.3               \\
{[}MW+Sgr+LMC{]}              & 0.8                   & 0.8                & -0.02        & -7.5        & 84.2         & 22.3                
\end{tabular}
\label{sat-pos}
\end{table*}

\subsection{The satellites}\label{dwarf}
The satellite galaxy models (Sgr and the LMC) are constructed directly following \cite{Vasiliev+21}. In brief, Sgr consists of a stellar component following a King profile \citep{King62} embedded in a dark matter spheroid. The LMC is modelled as a single spheroid with no distinction between stellar and dark matter. This work is only concerned with the effect on the Milky Way's disc, and a large spherical perturber on its first infall is sufficient to capture the disc response. Thus, the focus in modelling was given to reproducing the orbital configuration of satellites rather than their internal structure. In particlar, precisely replicating the state of the LMC is beyond the scope of this work, since its present day structure is shaped by its interaction with the Small Magellanic Cloud. \citep[e.g.][]{Choi+22,JimenezArranz+24}. See Table \ref{ncomps} for the satellite parameters and particle resolution.

\subsection{Running the simulations}\label{run-sim}

All models were evolved with Bonsai \citep{Bonsai,Bonsai-242bil}, a GPU accelerated $N$-body tree code, for a total of 3 Gyr using a smoothing length of 10 pc and an opening angle $\theta_{\mathrm{o}}=0.4$ radians. A total of 310 snapshots were initially generated per model, with a cadence of 9.77 million years. For models [MW+LMC], [MW+Sgr] and [MW+Sgr+LMC], we evolved the satellites alongside the MW, and set them up for infall using the solutions presented in \cite{Vasiliev+21}. 

For the [MW], [MW+LMC] and [MW+Sgr+LMC] models we use snapshot 289 occurring at $t=2.836$ Gyr as the present day, where the satellite positions best match the observed values. For the [MW+Sgr] model, the lack of the LMC changes the orbit of Sgr, such that the best fit was found between snapshots. Thus we produced an extra snapshot for Model [MW+Sgr] which occurs slightly later at $t=2.904$ Gyr. Table \ref{sat-pos} describes the initial and final position and velocity of the two satellites. 

The combined [MW+Sgr+LMC] model has both satellites `present day' position within a kpc of the estimated position from $Gaia$ which we consider a good recovery. The majority of the kinematics of the `present day' systems are within the observed errors, with the exception of the vertical motion of the LMC, which the model consistently underestimated in our tests, and the `y' velocity of Sgr which is too large in our model. For both the positions and velocities, the [MW+Sgr] and [MW+LMC] `one satellite' models are slightly worse fits because the orbit was determined in the presence of all three systems.

We also note that the influence of both satellites causes reflex motion and a change in the rotation axis of the disc. In the subsequent analysis we have re-centered and realigned the disc to the $z$ axis of angular momentum, $L_z$, of the inner 5 kpc of the disc in the `present day' snapshots. No azimuthal rotations were applied post-merger such that the orientation of the disc and the satellites are self-consistent.

Figure \ref{fig:a} shows the orbit of Sgr (top panel) and the LMC (bottom panel), for the simulations [MW+LMC] \& [MW+Sgr] (solid lines), and the combined simulation [MW+Sgr+LMC]  containing both (dashed lines). In the case of model [MW+Sgr] and [MW+Sgr+LMC], Sgr has completed approximately 2.5 orbits, and for models [MW+LMC] and [MW+Sgr+LMC], the LMC is on its first infall, having just passed its first pericentre. 

The bottom panel shows that the relatively low mass of Sgr in this model has very little effect on the orbit of the LMC over the last 3 Gyr, with only a minor deviation from the orbit of the LMC alone in Model [MW+LMC]. In contrast, the last 500 Myr of the orbit of Sgr is significantly affected, as shown in \cite{Vasiliev+21}. 

A detailed analysis of the differences in the orbit of Sgr and the resulting stellar stream is not the focus of the work, we instead focus on the impact of the satellites on the stellar disc. However, note that the streams from the [MW+Sgr] and [MW+Sgr+LMC] simulations are examined in \cite{Cunningham+23}, where they are used to validate the detection of a metallicity gradient with latitude in the Sgr stream in the $Gaia$ data, suggesting an initial radial metallicity gradient in the Sgr
dwarf galaxy of approximately $-0.1$ to $-0.2$ dex kpc$^{-1}$. 


\section{Results}
In this section, we analyse the disc response to the LMC and Sgr, separately and together, in contrast to the isolated disc. Specifically, we investigate and compare the morphology, mean radial velocity and mean vertical velocity in all four models. The morphology provides physical insight for the reader, while the radial and vertical velocity fields are potentially observable with current and next-generation surveys. Sections \ref{sec:morph} to \ref{sec:vz}, and Section \ref{sec:overtime} explores the evolution of the response over time. Through this approach, we aim to provide insights into the contrasting and combined roles of the recent interactions with the LMC and Sgr in shaping the morphology and kinematics of the Milky Way's disc. 


 \begin{figure}
    \includegraphics[width=\linewidth]{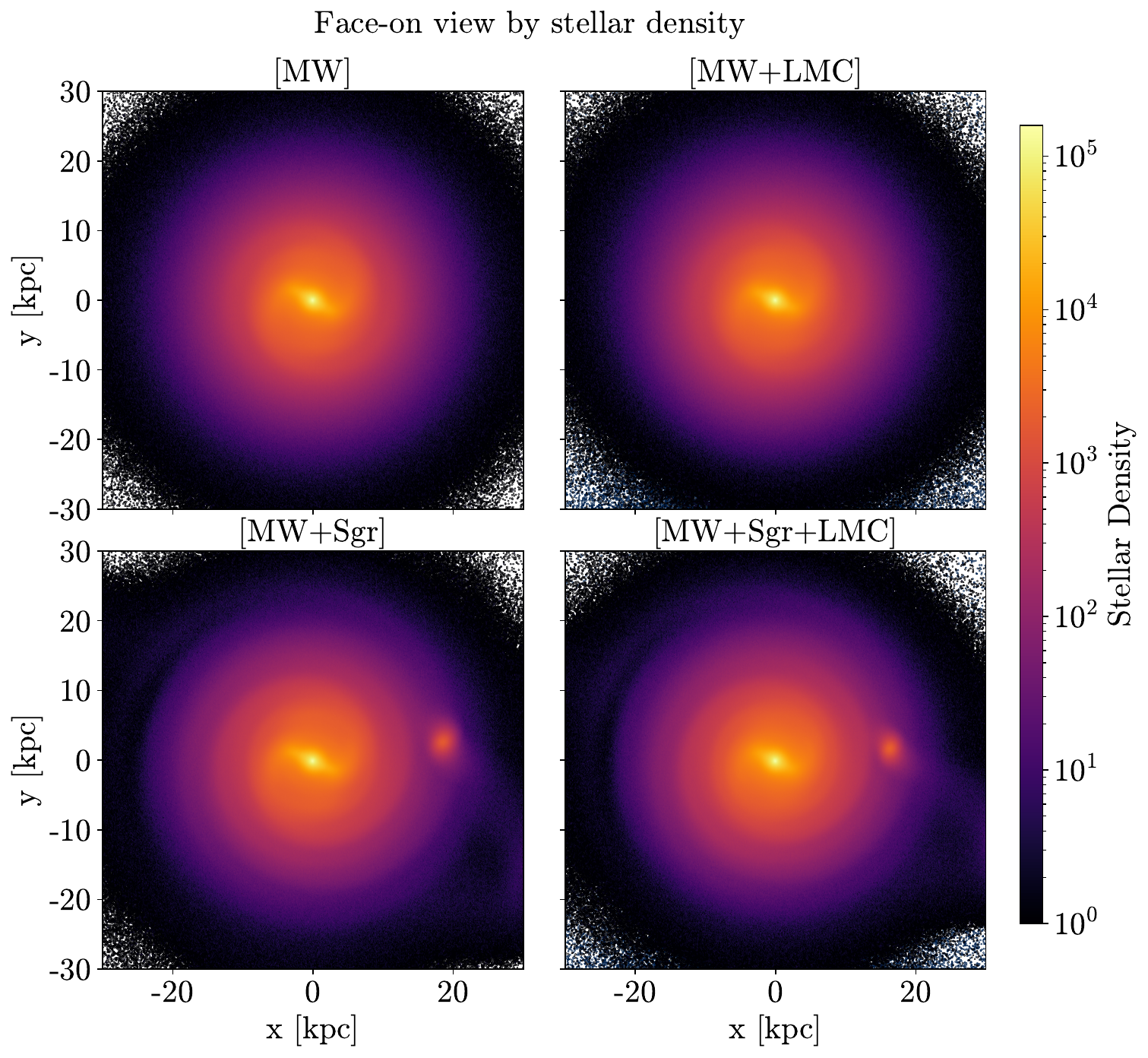}
    \caption{Density maps of the stellar disc in the $x-y$ plane at the present day for all four models. Overall, the disc presents a noticeable bar and spiral structure. Sgr is visible in models [MW+Sgr] \& [MW+Sgr+LMC] in the right hand part of the figure.
}\label{fig:face-on}
  \end{figure}

 \begin{figure}
    \includegraphics[width=\linewidth]{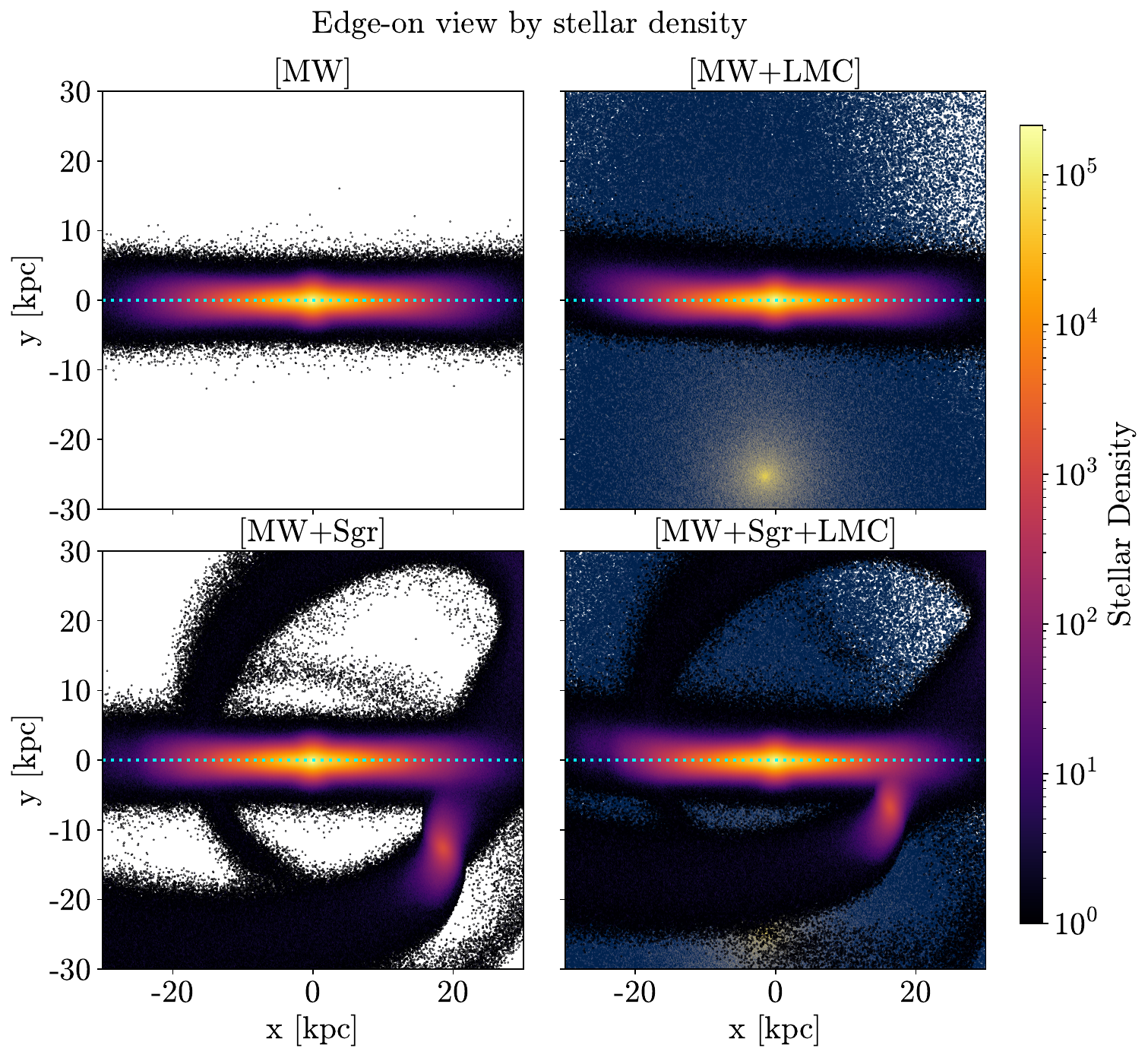}
    \caption{Density maps of the stellar disc, Sgr stream, and the LMC (in a separate color map) in the $x--z$ plane at the present day for all four models. The $z = 0$ plane is marked by a dotted cyan line. In the case of models including the LMC, [MW+LMC] \& [MW+Sgr+LMC], we can observe a warp of the disc, most strongly at negative $x$. The Sgr stream is present in the corresponding models [MW+Sgr] \& [MW+Sgr+LMC].}

\label{fig:edge-on}
  \end{figure}

\subsection{Large-Scale Morphology}\label{sec:morph}
Figure \ref{fig:face-on} shows the face on $(x-y)$ logarithmic stellar density maps of the present day snapshots of the four models; The isolated [MW] (top left), [MW+LMC] (top right), [MW+Sgr] (bottom left) and [MW+Sgr+LMC] (bottom right). The Sgr remnant is visible on the right hand side of panels [MW+Sgr] \& [MW+Sgr+LMC].

The top left panel shows that the unperturbed host galaxy [MW] contains a central bar with half-length $R_{\mathrm{b}}\approx4$ kpc. Just outside the bar region in Model [MW] is a faint two-armed bar driven spiral pattern, but the rest of the disc remains featureless after 3 Gyr owing to the dynamical heat of the disc, as intended.

The [MW+LMC], [MW+Sgr] \& [MW+Sgr+LMC] models all show the same bar and slight bar-driven spiral pattern, and the angle appears qualitatively preserved. While it is known that the impact of a satellite can induce a bar instability \citep[e.g.][]{Purcell+11}, in our models the bar instability exists in the host prior to interaction with the satellites, which then have only a minor effect on the bar evolution (see Section \ref{sec:fourier-vR} for further discussion. This is in contrast with the previous study of \cite{Laporte+18b_Sgr+LMC}, where their Sgr dwarf galaxy induces the bar instability. In this work we choose to have a preexisting bar such that the bar signature can be decoupled from the direct influence of the two satellites on the disc. Whether an earlier impact of Sgr on the Milky Way could have induced the bar remains an open question.

There is no immediately obvious effect on the face on disc morphology from the LMC in Model [MW+LMC], but models [MW+Sgr] \& [MW+Sgr+LMC] show that the repeated passage of Sgr have excited fine spiral structure in the Solar neighborhood and outer disc compared to the otherwise smooth disc in Models [MW] \& [MW+LMC] \citep[which is consistent with][]{Laporte+18b_Sgr+LMC}. 

Figure \ref{fig:edge-on} shows the same as Figure \ref{fig:face-on}, but for the edge-on $(x-z)$ plane, with the dashed cyan line marking the $z=0$ line. The LMC induces a clear warp in the outer disc of the [MW+LMC] \& [MW+Sgr+LMC] models, most clearly at negative $x$. As discussed in the introduction, it has long been known that the LMC could cause the warp in the Milky Way, and this is as expected from other models \citep[e.g.][]{Weinberg+Blitz06,Laporte+18a_LMC}, although see also \cite{Han+23_tiltwarp} for the contribution of a tilted halo. The [MW+Sgr] \& [MW+Sgr+LMC] models show the Sgr stream following the disruption of the dwarf, which is very similar to the models of \cite{Vasiliev+21} from which the initial conditions were adapted. The stream is examined further in \cite{Cunningham+23} but it is not the focus of this work.


\subsection{Radial velocity signatures}\label{sec:vR}
We also expect Sgr and the LMC each to leave a significant imprint in the kinematics of stars in the Milky Way's disc such as the radial velocities, $v_{\mathrm{R}}$. 

\subsubsection{Large scale kinematics}\label{sec:vR-largeScale}
\begin{figure}
    \includegraphics[width=\linewidth]{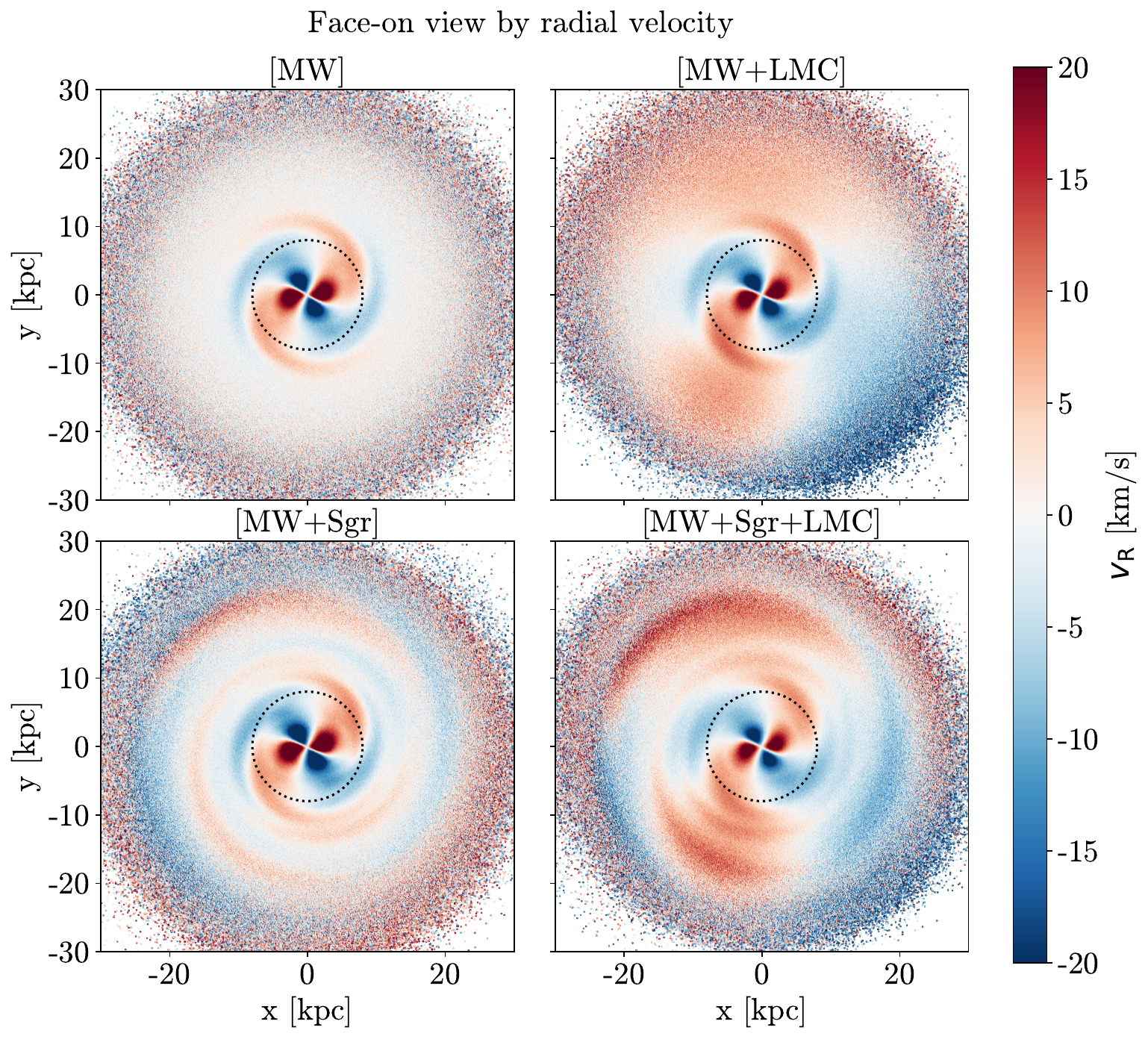}
    \caption{Face-on view of the Milky Way's stellar disc in the XY plane, at the present day for all four models, coloured by mean radial velocity. We notice the presence of a quadrupole in radial velocity in all models.}
    \label{fig:face-vR}
  \end{figure}

Figure \ref{fig:face-vR} shows the face on $x-y$ mean radial velocity, $\bar v_{\mathrm{R}}$  (km s$^{-1}$), maps for the present day snapshots of the [MW] (top left), [MW+LMC] (top right), [MW+Sgr] (bottom left) and [MW+Sgr+LMC] (bottom right) models. Note that while Figure \ref{fig:face-on} contained the satellite particles, \mvr\ maps show only disc particles. Model [MW] (the unperturbed host galaxy) shows a clear quadrupole in $\bar v_{\mathrm{R}}$ in the bar region, and a bar driven spiral pattern just outside the bar radius. Outside the Solar radius it appears otherwise homogeneous.

The presence of such a quadrupole structure is expected for a barred galaxy, and it has been observed and studied in the Milky Way \citep[e.g.][]{BLHM+19,GaiaCollab+23_mapasym,Leung+23,Hey+23}. This is not a new result, and we are not attempting to fit the structure or kinematics of the Milky Way bar with these models, but it is useful to compare the bar kinematics in the [MW] model with the [MW+LMC], [MW+Sgr] \& [MW+Sgr+LMC] models such that we can separate them from the satellite signatures. However, we do note that while the angle of the \mvr\ quadrupole appears to be preserved between models, it also appears to be qualitatively stronger in the [MW+Sgr] model than the isolated host Model [MW], and that it appears qualitatively weaker in the [MW+Sgr+LMC] model. We will return to this in Section \ref{sec:fourier-vR}.

Outside the bar region, the influence of each satellite on $\bar v_{\mathrm{R}}$ is clear, and distinct. At the present day the LMC has induced a large \mvr\ $m=1$ mode in the disc about the $y=0$ axis, with a secondary response at $y<0$ in the [MW+LMC] model.

In contrast, Sgr has induced multiple thin spiral like features in \mvr\ in the [MW+Sgr] model, kinematic counterparts to the spiral pattern seen in the density in Figure \ref{fig:face-on}. The \mvr\ field in the [MW+Sgr+LMC] model, which includes both the LMC and Sgr, is a clear composite of the signatures in the [MW+LMC] and [MW+Sgr] models. This combination is particularly striking, suggesting that we may be able to decompose the origin of large scale \mvr\ patterns in the Milky Way's disc, which are starting to be observed by large stellar surveys such as $Gaia$ and APOGEE, and learn about the nature of these interactions.

In general, we would not expect the small scale local \mvr\ pattern in the Milky Way Solar neighborhood \citep[e.g.][]{GaiaCollab+23_mapasym} to match the simulations, owing to secular dynamics in the disc such as non-satellite-induced spiral arms, which can change rapidly over 100s of Myr \citep[e.g.][]{Vislosky+23}. However, we would expect the large scale patterns to remain visible (albeit not as clearly) once we can resolve a larger fraction of the Milky Way disc. 

For example, \cite{Eilers+20} find a spiral-like band of outwards moving stars ($\sim5$ to 10 km $s^{-1}$) around $-15<x<-10$ kpc, which changes to an inwards moving band outside $x<-15$ kpc (see their Figure 1, speculatively linked to a Milky Way spiral arm). This is qualitatively consistent with the [MW+Sgr] and [MW+Sgr+LMC] models which include Sgr, where the transition from outwards moving (red) to inwards moving (blue) stars happens at $x\approx-15$ kpc, and with similar amplitudes. Thus, we suggest that the large scale radial velocity feature from \cite{Eilers+20} may be a tidally induced spiral caused by the recent influence of the Sgr dwarf galaxy on the Milky Way disc. The inwards and outwards flow patterns are an expected signature of spiral structure and correlation does not necessarily imply causation. However, the location of the tidal arm is sensitive to the disc crossing time of Sgr, and we note that the disc crossing of Sgr in our model \citep[using the orbit of][]{Vasiliev+21} naturally puts this feature in the right location and amplitude to be consistent with the data. In our models, the outwards moving band is comprised of two separate features, which is not the case in \cite{Eilers+20}, but the resolution of the data is likely insufficient to resolve this, if it is the case. Data from future surveys will be needed to confirm or refute the connection.

Unfortunately, the stronger signal of the LMC's effect is not visible in current data sets. The \mvr\ fields in the Solar neighbourhood is close to 0 in the [MW+LMC] model, with the stronger \mvr\ signals occurring further from the Solar azimuth. However, we predict a large scale radial velocity feature to be observed in future surveys of the disc such as SDSS-V Milky Way Mapper \citep{Kollmeier+19}.

\subsubsection{Fourier decomposition of \mvr}\label{sec:fourier-vR}

For a more quantitative analysis, we employ Fourier decomposition on the radial velocity signal as a function of Galactocentric radius. 
This is done by dividing the particle data into 0.5 kpc radial bins. Within each radial ring, we calculate the mean radial velocity azimuthal bins and perform a Fourier transform on the resulting distribution. The number of azimuthal bins is dynamically adjusted based on the particle count in each radial bin.

\begin{figure}
    \includegraphics[width=\linewidth]{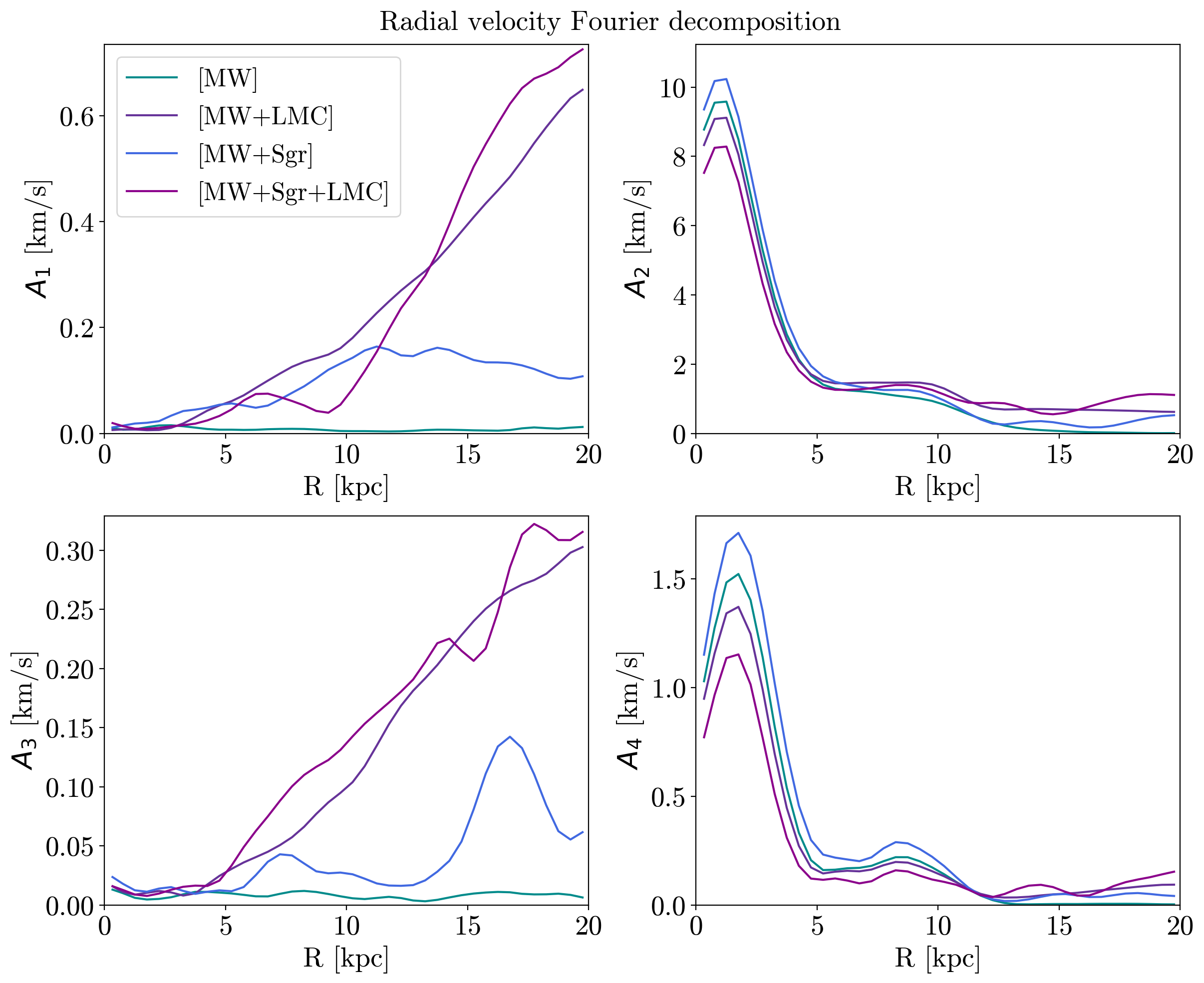}
    \caption{Amplitude, $A$, of the $m = 1$ (upper left), $2$ (upper right), $3$ (lower left) and $4$ (lower right) modes from the Fourier decomposition of the radial velocity field as a function of Galactocentric radius.
}\label{fig:pres-ff}
  \end{figure}

Figure \ref{fig:pres-ff} shows the amplitude $A$ of the first 4 Fourier components from the decomposition of the mean radial velocity, \mvr, for the four models as a function of Galactocentric radius. Our control model, the isolated disc ([MW]; green) shows negligible signal (consistent with noise) in the odd $m = 1$ and $m = 3$ modes, with corresponding amplitudes  $A_1$ \& $A_3$. Conversely, the even $m = 2$ and $m = 4$ components with corresponding amplitudes $A_2$ \& $A_4$, have a strong peak in the inner galaxy. The $m = 2$ mode shows the quadrupole of the bar, and the $m = 4$ mode shows a higher-order bar structure (octopole). The tail between $R\approx5-12.5$ kpc shows the bar-driven spiral structure in the otherwise quiet disc. 

\begin{figure}
    \includegraphics[width=\linewidth]{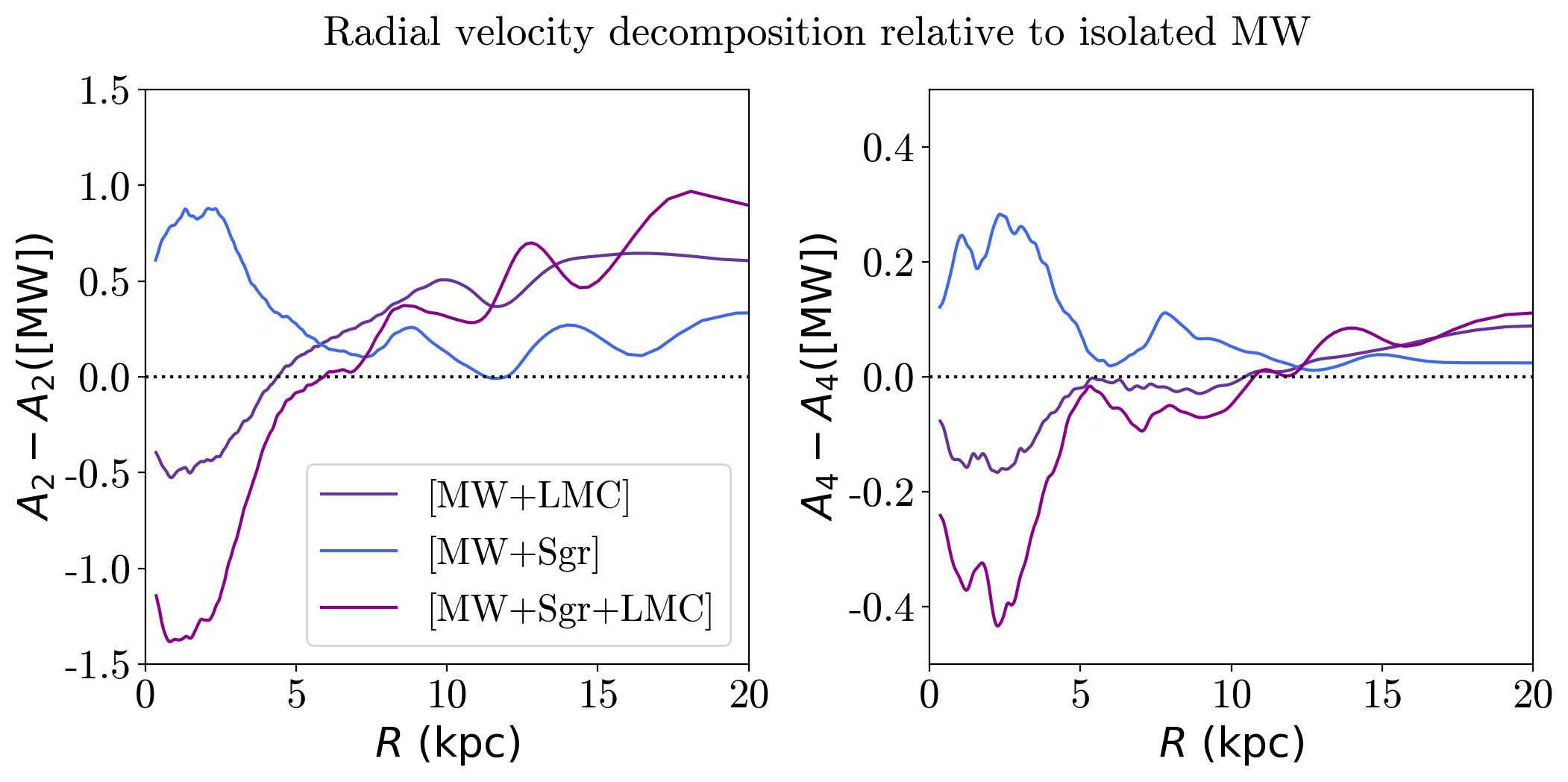}
    \caption{Amplitude of $m = 2$ (left) and $m = 4$ (right) modes from the Fourier decomposition of the radial velocity field as a function of Galactocentric radius for the three merger models after subtracting the power from the isolated host model.
}\label{subtract}
  \end{figure}
The left column shows that the odd Fourier modes are strongly excited by the LMC. The [MW+LMC] model shows a steady growth in both $A_1$ \& $A_3$ from around $R\approx3$ kpc out to the outer disc, their combination making up the asymmetric dipole seen in the top right panel of Figure \ref{fig:face-vR}. The [MW+Sgr] model shows that Sgr also induces both $m=1$ and $m=3$ modes in \mvr, albeit with a significantly lower amplitude. Interestingly, in the combined model, the $A_1$ amplitude appears to be stronger in the outer disc, than the LMC alone, yet weaker in the inner disc, suggesting constructive and destructive interference respectively. The $m = 3$ modes are constructive across all $R$, expect for a small spike around $R=15$ kpc. 

The findings from Figure \ref{fig:face-on} with respect to the bar are also quantitatively represented in the case of the amplitudes of the $m = 2$ \& $m = 4$ modes. The strength of the quadrupole (and the octopole) in the inner galaxy varies with interaction with the satellites. The [MW+Sgr] model shows an increase in the even modes of approximately $10\%$, while the [MW+LMC] model shows a decrease of approximately $10\%$. We speculate that this may be owing to the phase of the bar with respect to the satellite orbit, where the torque from the satellite increases or decreases its angular momentum. Interestingly, the combined [MW+Sgr+LMC] model shows the lowest amplitude, showing a non-linear combination of effects from the two satellites. However, bar dynamics are not the focus of this paper and we defer a more thorough exploration to future work (Petersen et al. in prep).

In Figure \ref{fig:pres-ff}, we can see that the even Fourier components are dominated by the bar within 10 kpc in our models. However, Models [MW+LMC] and [MW+Sgr+LMC] show $m = 2$ \& $m = 4$ modes of comparable strength to the odd modes for those models in the left column ($A_2\sim0.5$, $A_4\sim0.1$). In each case, the LMC induces the stronger $m = 2$ \& $m = 4$ modes, and the contribution of Sgr is constructive and destructive at different radii. 

To more closely examine the effects of the satellites outside the bar region, Figure \ref{subtract} shows the amplitude of the $m=2$ and $m=4$ Fourier modes of the mean radial velocity field for the three merger models after subtracting the amplitude of the respective mode from the isolated host. Figure \ref{subtract} shows that the LMC causes a net decrease in the bar amplitude ($R<5$ kpc), but an increase in the strength of $A_2$ across the disc outside the bar region, and of $A_4$ in the outer disc. In contrast, Sgr causes an enhancement to the bar strength, but only a slight net increase in $A_2$ outside the bar region, but with a oscillatory pattern with radius. Sgr also causes a larger increase in $A_4$ around the Solar radius than the LMC, which has little effect inside 10 kpc. 

Interestingly, the model with both satellites shows a combination of constructive and destructive interference in both $A_2$ and $A_4$. Most strikingly, the bar amplitude is lowest in the [MW+Sgr+LMC] model, despite Sgr alone causing an enhancement. Similarly, the enhancement in $A_4$ around the Solar radius in the [MW+Sgr] model, turns into a reduction in $A_4$ in the [MW+Sgr+LMC] model. 

\begin{figure}
    \includegraphics[width=\linewidth]{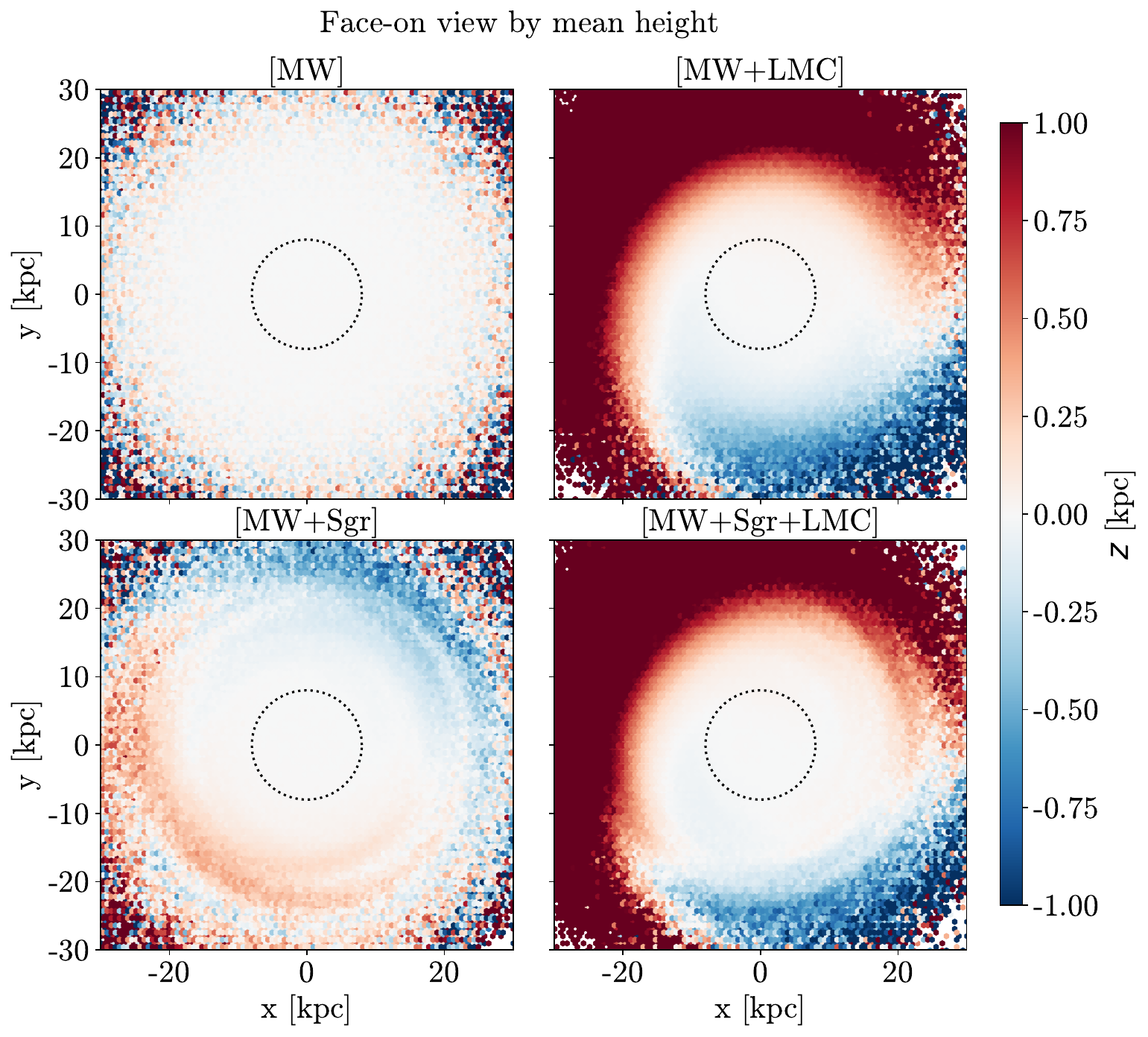}
    \caption{Face-on view of the Milky Way's stellar disc in the $x-y$ plane, at the present day for all four models, coloured by mean height, $\bar z$ (kpc).
}\label{fig:face-z}
  \end{figure}

  \begin{figure}
    \includegraphics[width=\linewidth]{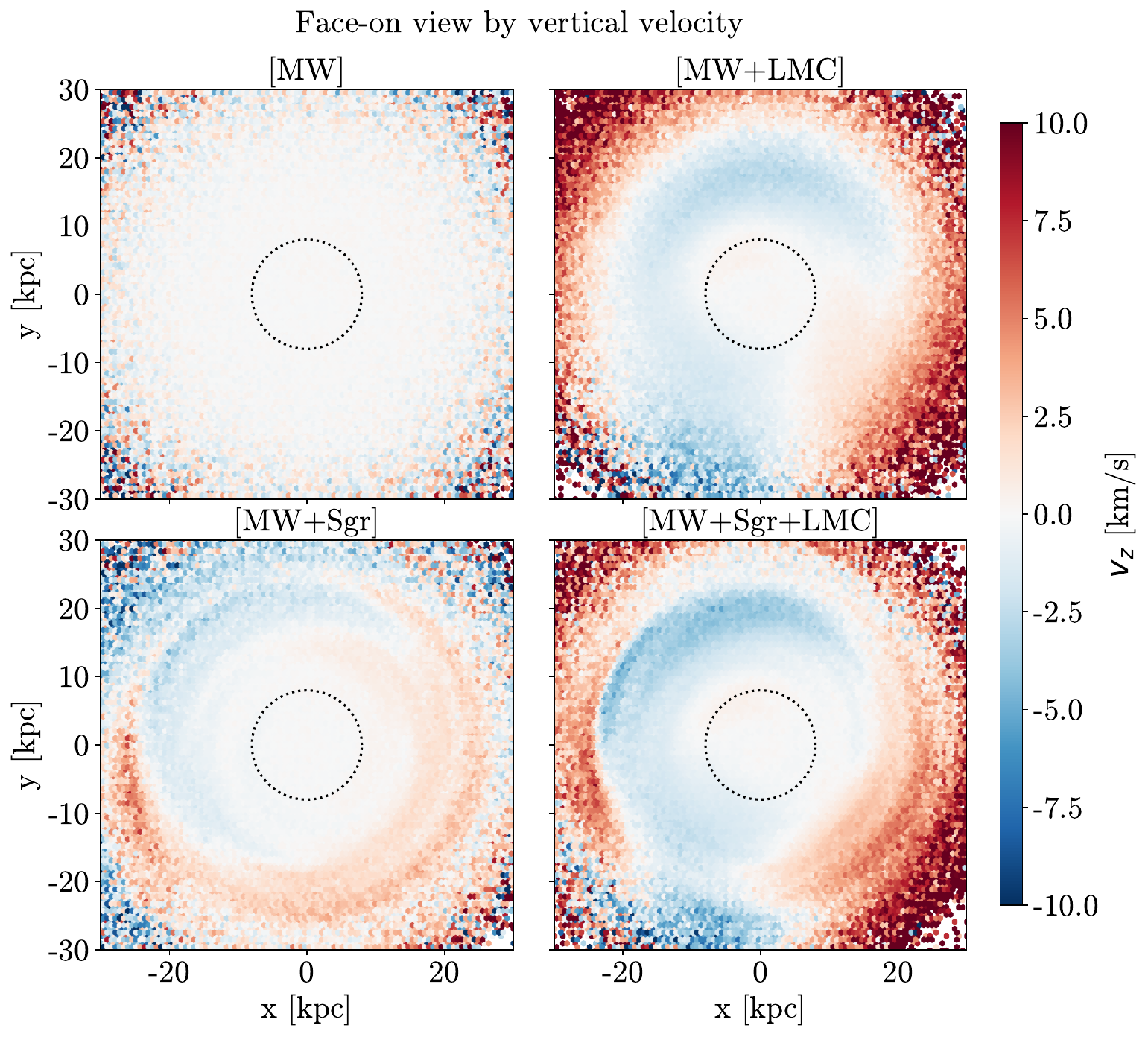}
    \caption{Face-on view of the Milky Way's stellar disc in the $x-y$ plane, at the present day for all four models, coloured by mean vertical velocity, $\bar v_z$ (km s$^{-1}$).
}\label{fig:face-vz}
  \end{figure}


\subsection{Vertical Structure and Kinematics}\label{sec:vz}
\subsubsection{Vertical morphology}
Figure \ref{fig:face-z} shows the face on maps of mean vertical position, $\bar z$ (kpc) for the four models. The isolated host galaxy, [MW], does not experience any vertical perturbation (e.g. the bar does not buckle), and the vertical structure (Figure \ref{fig:face-z}) remains smooth and featureless.

The top right panel of Figure \ref{fig:face-z} shows that the LMC (which is just past its its first pericentre) produces a clear warp in the disc of the [MW+LMC] model \citep[as expected from previous studies, e.g.][]{gomez2013MNRAS.429..159G,Laporte+18a_LMC} with slight more than half the outer disc having a positive $\bar z$. We allow the colorbar to saturate to resolve the weaker features from Sgr on the same scale in the [MW+Sgr] panel.

The specific configuration of our warp is very similar to earlier studies \citep[see e.g. Figure 6 of][which displays the same orientation of the warp]{Laporte+18a_LMC}. Note that we do not accurately recover the line of nodes which lies close to along the negative $x$, $y=0$ axis in the Milky Way, or the twisting or procession of the line of nodes \citep[see][]{Poggio+20,CabreraGadea+24,Jonsson+24}. As discussed in \cite{Laporte+18a_LMC}, this is very time dependent owing to the the precessing nature of the warp \citep[e.g.][]{Poggio+20} and the mismatch is unsurprising given the uncertainties in the Milky Way and LMC mass distribution and LMC orbit \citep[or it could be explained by a tilted halo, see][]{Han+23_tiltwarp}.

The lower left panel of Figure \ref{fig:face-z} shows that the last 3 Gyr of Sgr's orbit\footnote{Interactions between the Milky Way and Sgr prior to this may leave a large imprint on the disc, but the orbit and mass loss history of the actual Sgr dwarf galaxy are difficult to constrain over longer periods, and we examine here the effect of the orbital solutions of \cite{Vasiliev+21} on the disc.} also produces a slight warp in the disc of [MW+Sgr] model, although the angle of the warp is lagging approximately 90 deg behind the LMC induced warp in the [MW+LMC] model, which may be expected because the Sgr orbital plane is approximately 90 degrees from that of the LMC. Note that the warp is both significantly weaker, and in addition to the global dipole, numerous fine ripples can be seen similar to the $\bar v_{\mathrm{R}}$ maps in Section \ref{sec:vR}. The lower right panel shows that again the signatures are combined, where the LMC dominates the warp, while Sgr induces ripples on top of the warp, as found in \cite{Laporte+18b_Sgr+LMC}. 

\subsubsection{Vertical kinematics}
Figure \ref{fig:face-vz} shows the mean vertical motion, $\bar v_z$ (km s$^{-1}$) for the four models. As with Figure \ref{fig:face-z}, the isolated host shows no coherent signal in the vertical kinematics.

The upper right panel shows that the LMC induces a large scale signature in mean vertical velocity, with the majority of the disc outside $\sim20$ kpc moving upwards, and the majority of the disc between $10-20$ kpc moving downwards with respect to the galactic centre. The lower left panel shows that Sgr induces several weaker ripples, but also an overall dipole in the mean vertical velocity at a different angle to that of the LMC. Again, the lower right panel shows that the mean vertical velocity signature of the LMC and Sgr combined in the [MW+Sgr+LMC] model are a composite of the individual LMC and Sgr signatures from the [MW+LMC] \& [MW+Sgr] models. These results are also consistent with the previous results of less resolved simulations \citep[e.g. Fig's. 23-24 of][]{Laporte+18b_Sgr+LMC}, although the `warm disc' setup also allows us to separate the Sgr-induced features from the secular disc features.

\subsubsection{Fourier Decomposition of $\bar v_z$}\label{sec:fouriervz}
\begin{figure}
    \includegraphics[width=\linewidth]{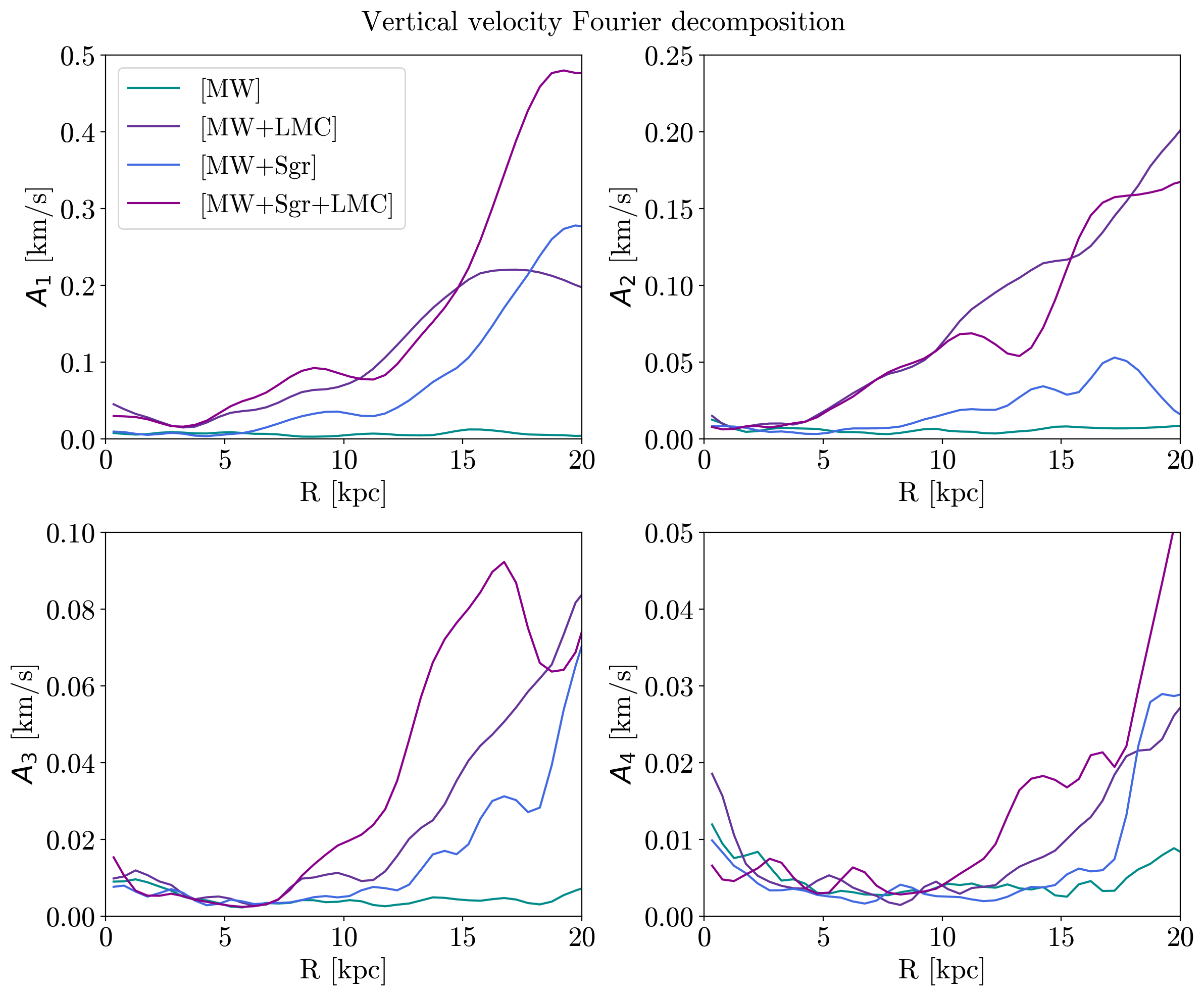}
    \caption{Amplitude, $A$, of the $m = 1$ (upper left), $2$ (upper right), $3$ (lower left) and $4$ (lower right) modes from the Fourier decomposition of the mean vertical velocity field as a function of Galactocentric radius.
}\label{fig:pd-z-ff}
  \end{figure}

We also perform a Fourier decomposition of the vertical velocity, following the method described in Section \ref{sec:fourier-vR}. Fig.\ref{fig:pd-z-ff} displays a quantitative representation of the satellites' influence on the disk. As observed in the morphology and mean vertical velocity field in Figure's \ref{fig:face-z} \& \ref{fig:face-vz} the isolated host has no coherent structure in the Fourier decomposition beyond the level of noise.

Interestingly, while the odd modes of the radial velocity decomposition ($m = 1$, $m = 3$) were dominated by the effect of the LMC outside the Solar radius (see Fig. \ref{fig:pres-ff}), the left column of Figure \ref{fig:pd-z-ff} shows that the contribution of the LMC and Sgr to the odd components of the vertical velocity decomposition are much closer to comparable, with the amplitude of $A_1$ \& $A_3$ in the outer disc from the LMC being only $0.8-2$ times that of Sgr. The right column shows that the LMC does dominate the amplitudes $A_2$ \& $A_4$ of the even modes ($m = 2$, $m = 4$), with only a minor contribution from Sgr. 

\subsection{Fourier mode amplitude over time}\label{sec:overtime}

  \begin{figure}
    \includegraphics[width=\linewidth]{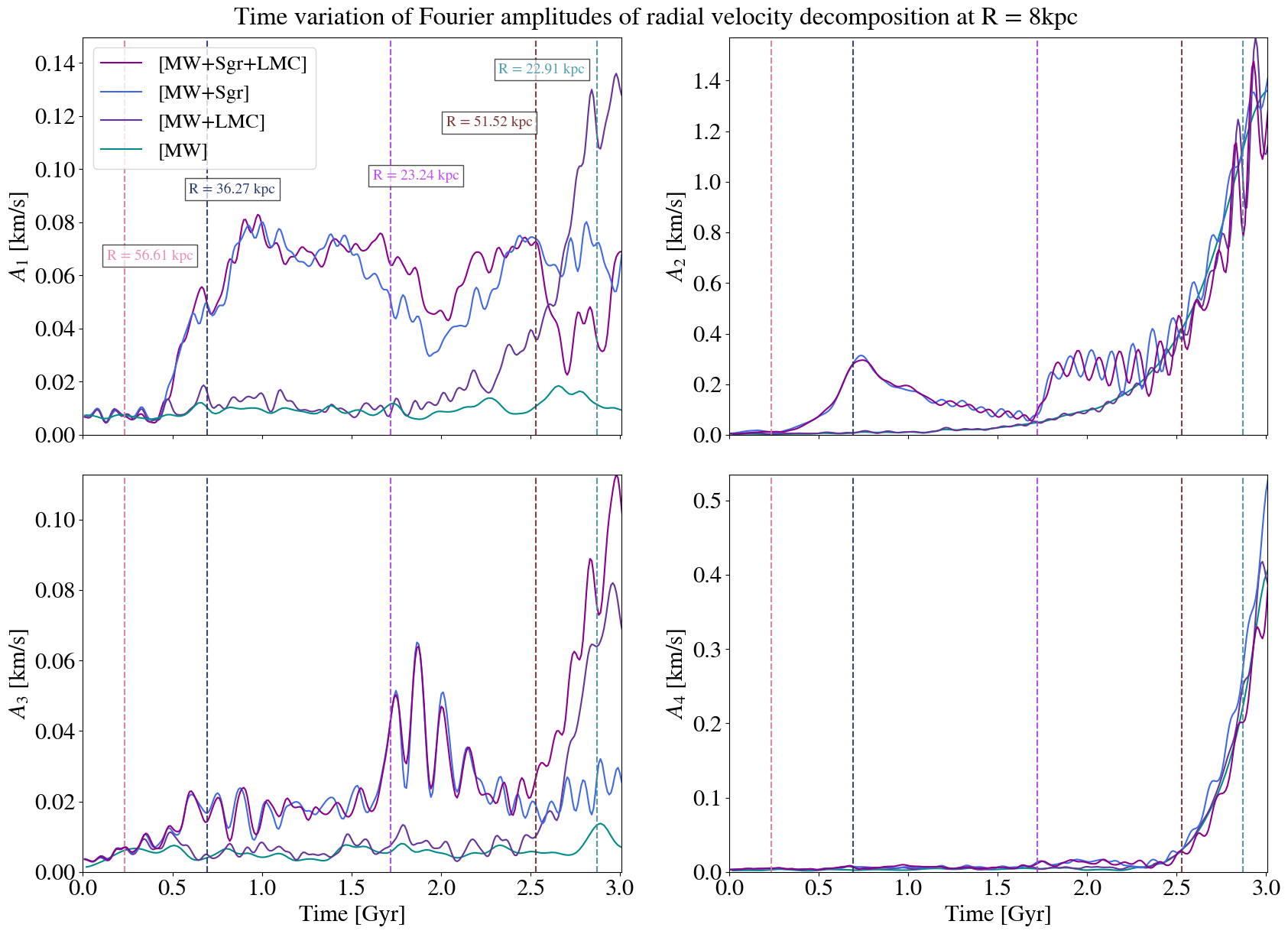}
    \caption{Amplitude of the first four Fourier components of the decomposition of the mean radial velocity field around an annulus of $R=8$ kpc as a function of time, smoothed with a one dimensional Gaussian filter with standard deviation $\sigma = 2$. The vertical lines mark the disc crossings of the Sgr dwarf, with the radii of each crossing labeled.
}\label{fig:time-ff}
  \end{figure}

  \begin{figure}
    \includegraphics[width=\linewidth]{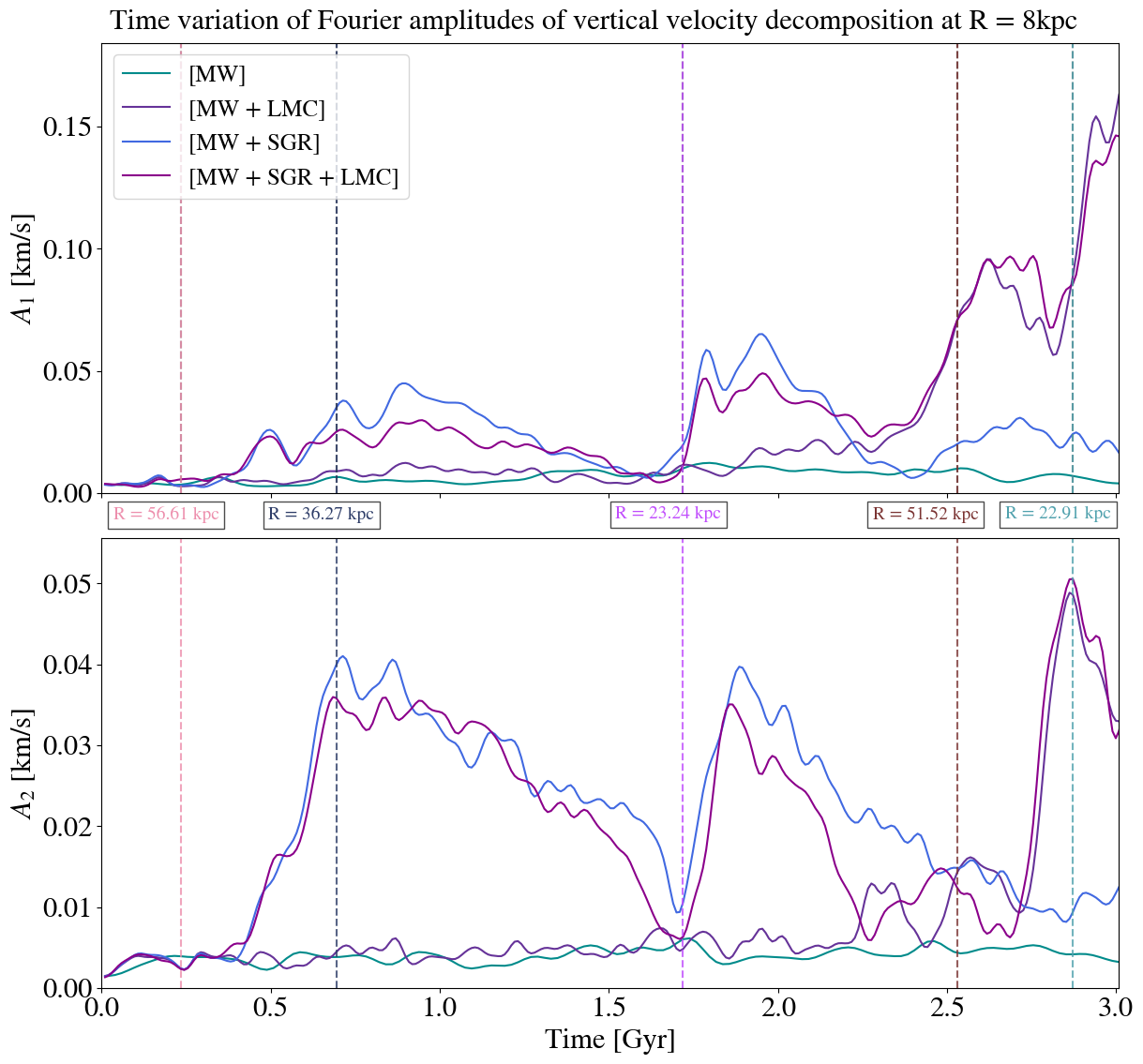}
    \caption{Amplitude of the first two Fourier components of the decomposition of the mean vertical velocity field around an annulus of $R=8$ kpc as a function of time, smoothed with a one dimensional Gaussian filter with standard deviation $\sigma = 2$. Note that $A_3$ and $A_4$ were indistinguishable from noise at $R=8$ kpc and have thus been omitted
.}\label{fig:vertime-ff}
  \end{figure}

The advantage of simulations is that we can also examine the evolution of the modes over time. In this section, we direct our attention to the temporal evolution of the amplitude of both radial and vertical Fourier decomposition modes ($m = 1, 2, 3, 4$) at a fixed radius ($R = 8$ kpc) corresponding to around the Solar radius. The methodology employed for amplitude calculations is the same as the approach outlined in Section \ref{sec:fourier-vR}. 

Figure \ref{fig:time-ff} shows the time evolution of $A_1$ to $A_4$, the amplitude of the first four Fourier components of the radial velocity decomposition in the four models. Note that the isolated host [MW] model contains no odd radial velocity modes, but the growth of the bar dominates the signal in the even $m=2$ and $m=4$ modes at late times at the Solar radius. We also note that while the growth of the $m=2$ and $m=4$ modes is smooth in the isolated [MW] model, the models that contain Sgr show periodic oscillations in the amplitudes $A_2$ and $A_4$. This occurs because of the overlap of the $m=2$ pattern induced by the galactic bar, and the $m=2$ patterns induced by the Sgr impacts. The pattern speeds of the bar and the induced wave are different, such that the modes show constructive and destructive interference. The oscillation in $A_2$ gets significantly stronger following the third disc crossing of the Sgr dwarf model at $R\sim23$ kpc.


The vertical dashed lines mark when Sgr crosses the disc in the [MW+Sgr+LMC] model, matching Figure \ref{fig:a}. The first passage of Sgr around 0.25 Gyr has little to no effect on the disc owing to the larger radius, yet the second passage at $R\sim36$ kpc at $\sim0.7$ Gyr induces a significant $m=1$ and $m=2$ mode in the radial velocities, and a small $m=3$ mode. The third disc crossing at around 1.75 Gyr decreases the $m=1$ mode, but increases the $m=2$ and $m=3$ modes. The fourth crossing gives a small increase to $m=1$ and $m=3$ but by then the influence of the LMC in the [MW+LMC] \& [MW+Sgr+LMC] models are dominant in the odd modes.

Figure \ref{fig:vertime-ff} shows the same as Figure \ref{fig:time-ff} but for the first two Fourier components of the mean vertical velocity. Note that we do not show $A_3$ or $A_4$ because they are indistinguishable from noise at $R=8$ kpc. Similar to the radial case, the second and third disc crossings of Sgr induce $m=1$ and $m=2$ modes in the disc, yet the fourth and fifth crossings of Sgr have little effect. The signal at late times is dominated by the LMC's presence in the  [MW+LMC] \& [MW+Sgr+LMC] models, yet the $m=1$ mode grows steadily from $\sim2.25$ Gyr as the LMC falls into the host, while the $m=2$ mode grows rapidly from $\sim2.75$ Gyr. There is a dip in $A_1$ corresponding to the peak in $A_2$ which occurs close to the pericentric passage of the LMC. 


\begin{figure}
    \includegraphics[width=\linewidth]{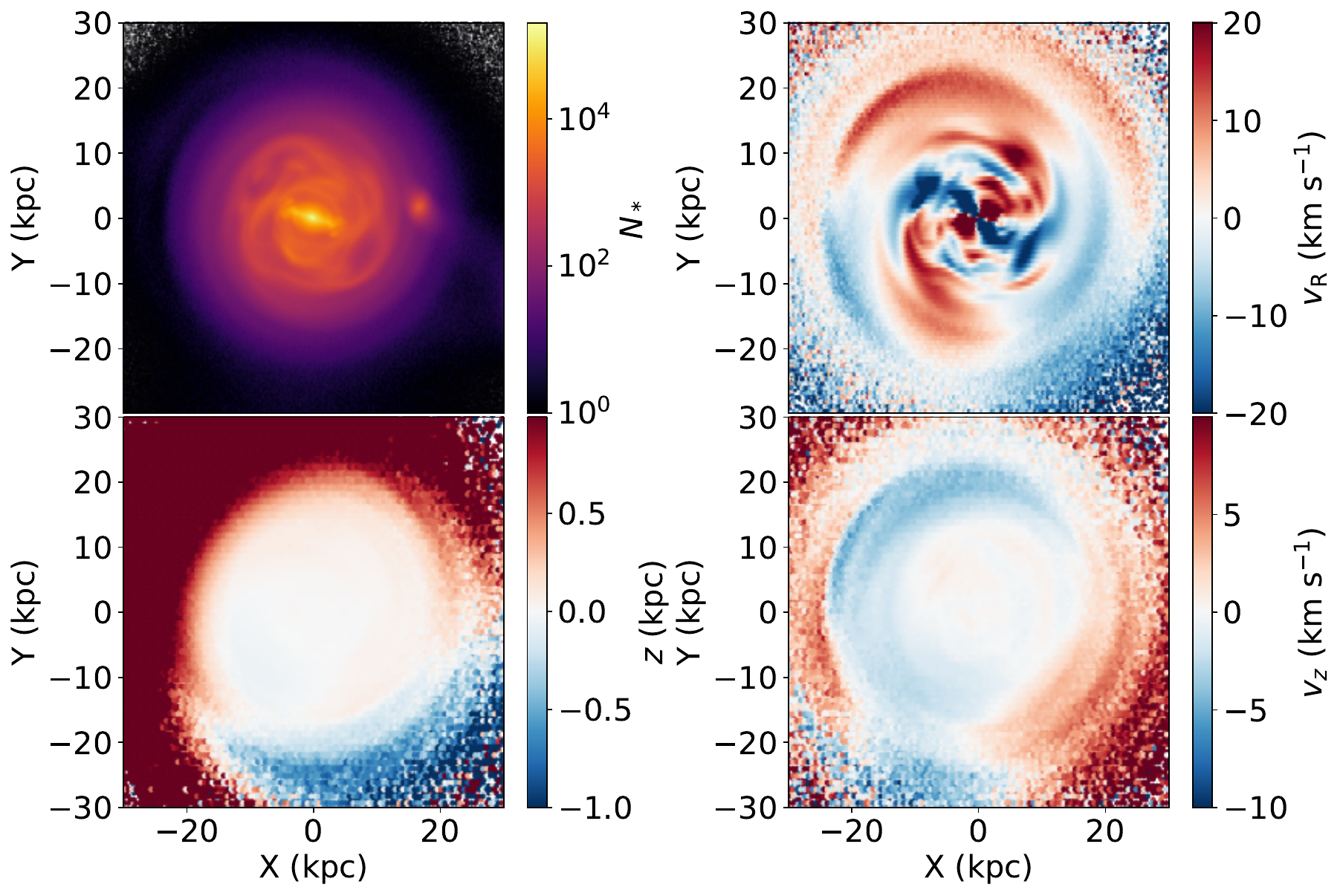}
    \caption{Surface density (upper left), mean radial velocity (upper right), mean height (lower left) and mean vertical velocity (lower right) for the [cMW+Sgr+LMC] model, which has the same satellite perturbers as the [MW+Sgr+LMC] model, but the host has a colder disc which is more susceptible to secular structure formation.
}\label{fig:colder}
\end{figure}

\section{The effect of the `hot' disc}\label{sec:cold-disc}
The models presented above assume a high velocity dispersion across the disc component in order to cleanly resolve the kinematic imprint of the satellites without secular structure (other than the central bar).

In this section we relax this restriction and present a fifth model which has the same orbital solutions as the [MW+Sgr+LMC] model above, i.e. including the LMC and Sgr, but using an isolated host with a colder disc [cMW+Sgr+LMC]. Note that we do not change the total mass of the host galaxy in the model, but we reduce the vertical scale height of the disc from 0.4 kpc to 0.28 kpc, and we reduce the scale radius of the disc radial velocity dispersion from 8 to 6 kpc. This maintains a realistic dispersion in the galactic centre, but more reasonable values in the Solar neighborhood and outer disc. 

Figure \ref{fig:colder} shows the surface density (upper left), mean radial velocity (upper right), mean height (lower left) and mean vertical velocity (lower right) for the colder disc model [cMW+Sgr+LMC] for comparison with the other figures. The upper left panel shows that the density shows more secular transient spiral structure within 15 kpc which was not present in the warmer disc models above, and which will leave its imprint in the kinematics of the disc. As such, the top right panel shows that the mean radial velocity field in the colder disc shows significantly more substructure than in Figure \ref{fig:face-vR} \citep[where the specific pattern will change over short timescales; e.g.][]{Vislosky+23}. However, the large scale influence of the two satellites on the $v_{\mathrm{R}}$ field in the outer disc remains clearly visible. 

Interestingly, the vertical warp and vertical velocity in the lower left and lower right panels look very similar to the [MW+Sgr+LMC] model in Figures \ref{fig:face-z} and \ref{fig:face-vz}, suggesting that the disc structure does not significantly alter the response over these timescales. However, the present day vertical dynamics of the disc are heavily influenced by the recent arrival of the LMC, which the disc is only beginning to respond to recently, leaving little time for that response to be influenced by the disc velocity dispersion. It is difficult to tell by eye whether the finer features induced by Sgr are affected by the change in the disc model, thus this will be discussed further in the next Section using the Fourier decomposition.

We also note here that neither the warm disc model [MW] or the cold disc model [cMW] undergo a buckling event by the `present day' snapshots of the simulations, and as such neither model has any significant vertical dynamical signatures from the isolated disc without the influence of the satellites.

\subsection{Fourier decomposition of the colder disc model}\label{sec:cold_fourier}

\begin{figure}
    \includegraphics[width=\linewidth]{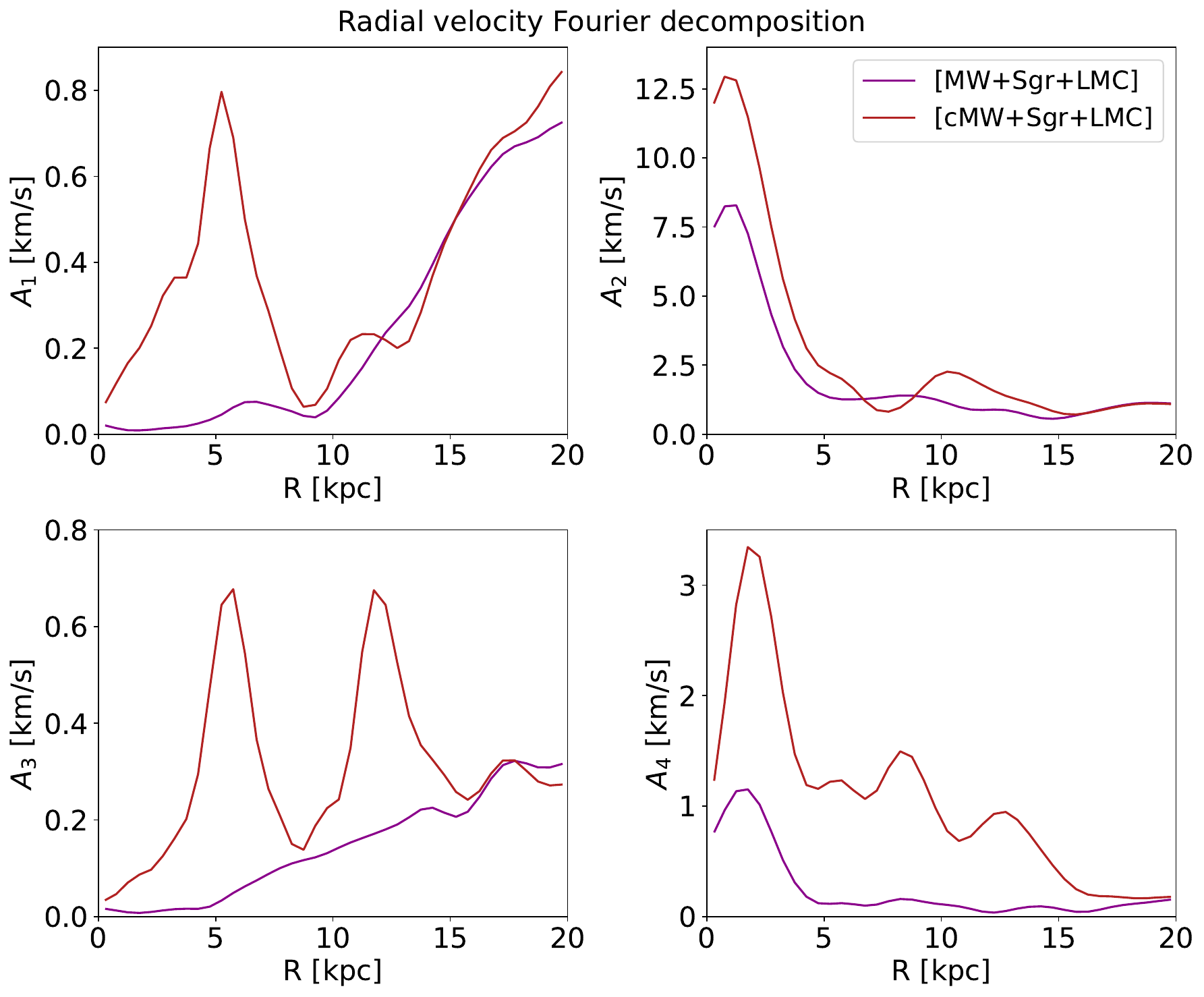}
    \caption{Same as Figure \ref{fig:pres-ff} but comparing Model [MW+Sgr+LMC] with the colder host disc in Model [cMW+Sgr+LMC].}
    \label{fig:cold_rad_pres}
  \end{figure}

\begin{figure}
    \centering
    \includegraphics[width=\linewidth]{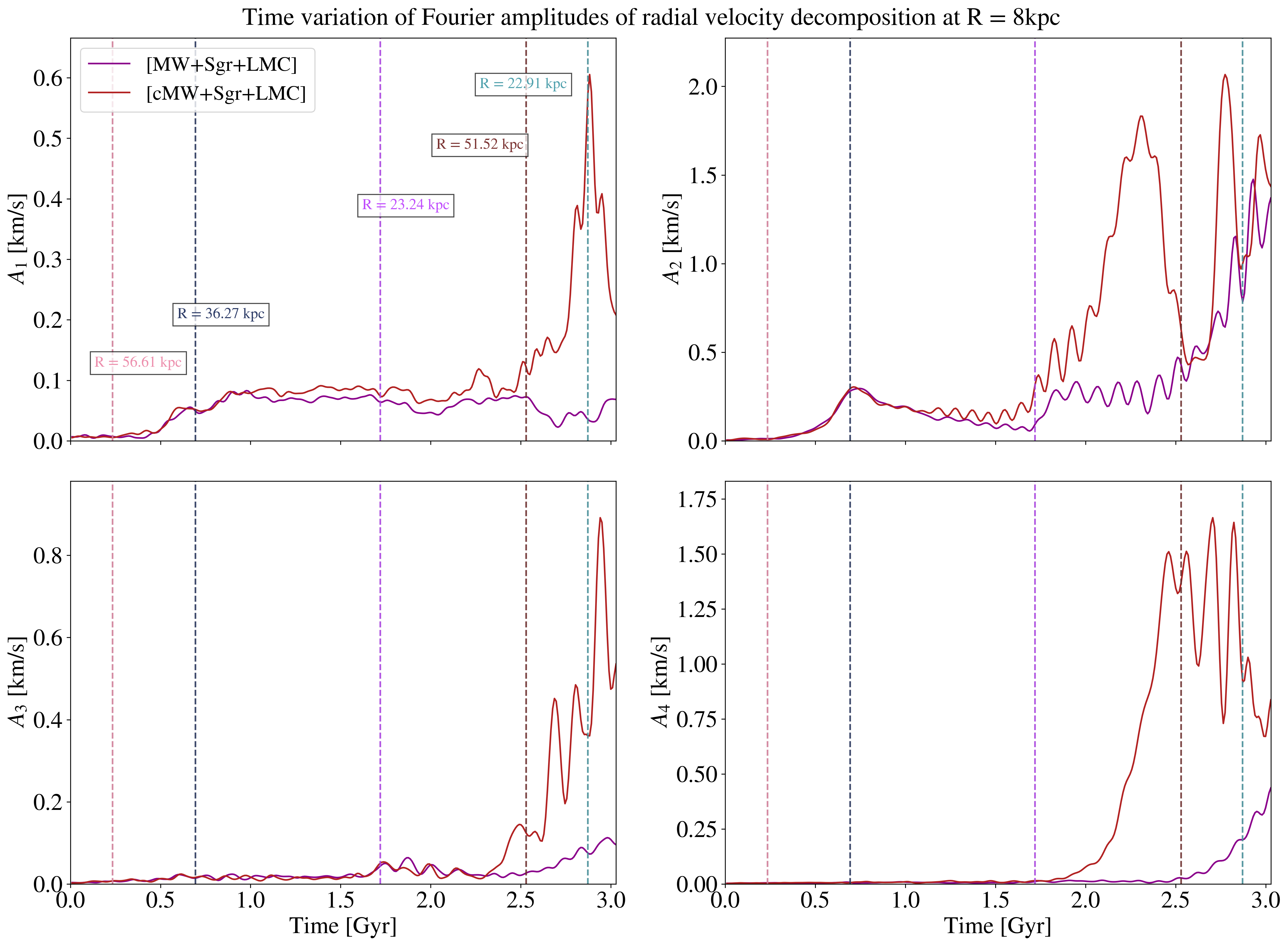}
    \caption{Same as Figure \ref{fig:time-ff} but comparing Model [MW+Sgr+LMC] with the colder host disc in Model [cMW+Sgr+LMC].}
    \label{fig:cold_rad_time}
\end{figure}

Figure \ref{fig:cold_rad_pres} shows the Fourier decomposition of the `present day' mean radial velocity fields for Model [cMW+Sgr+LMC] (red) compared to Model [MW+Sgr+LMC] (purple). $A_2$ and $A_4$ show that the bar is stronger in the cold disc as we would expect, and there are also higher amplitude fluctuations in all four Fourier components than in the warm disc model. We note that the $m=1$ signature induced by the LMC is very clearly visible in the outer disc, and the slow increase in the $m=3$ mode from the LMC is likely separable from the local fluctuations in the outer disc, but may be dominated by secular dynamics within 15 kpc or so.

Figure \ref{fig:cold_rad_time} shows the Fourier decomposition of the mean radial velocity fields at $R\sim8$ kpc over time for Model [cMW+Sgr+LMC] compared to Model [MW+Sgr+LMC]. While the modes induced by the second disc crossing of Sgr at $t\sim0.7$ Gyr, $R\sim36$ kpc remain similar, the $m=1$ and $m=2$ modes mix away slightly faster in the warm model that the cold disc model as expected, retaining higher amplitudes prior to the third disc crossing. By the present day the amplitude of the odd modes, $A_1$ and $A_3$, are significantly higher in the cold disc model than in the warm disc model following the fourth impact of Sgr. Unlike the warm disc model, $A_2$ and $A_4$ are of comparable amplitudes at the later stages of the model. They also show significant oscillation, where the even modes are dominated by $m=2$ or $m=4$ at different times.

This has important implications for the observations: We do not expect to find perfect, clean, kinematic signatures in the radial velocities such as seen in our warm disc model in the real Milky Way. Thus, on local scales such as the Solar neighborhood the large scale pattern may be obscured by secular dynamical features. However, future surveys which cover a large fraction of the disc should be able to resolve the large scale kinematic imprint of both Sgr and the LMC on top of small scale fluctuations, allowing us to constrain their mass and orbit.

\begin{figure}
    \includegraphics[width=\linewidth]{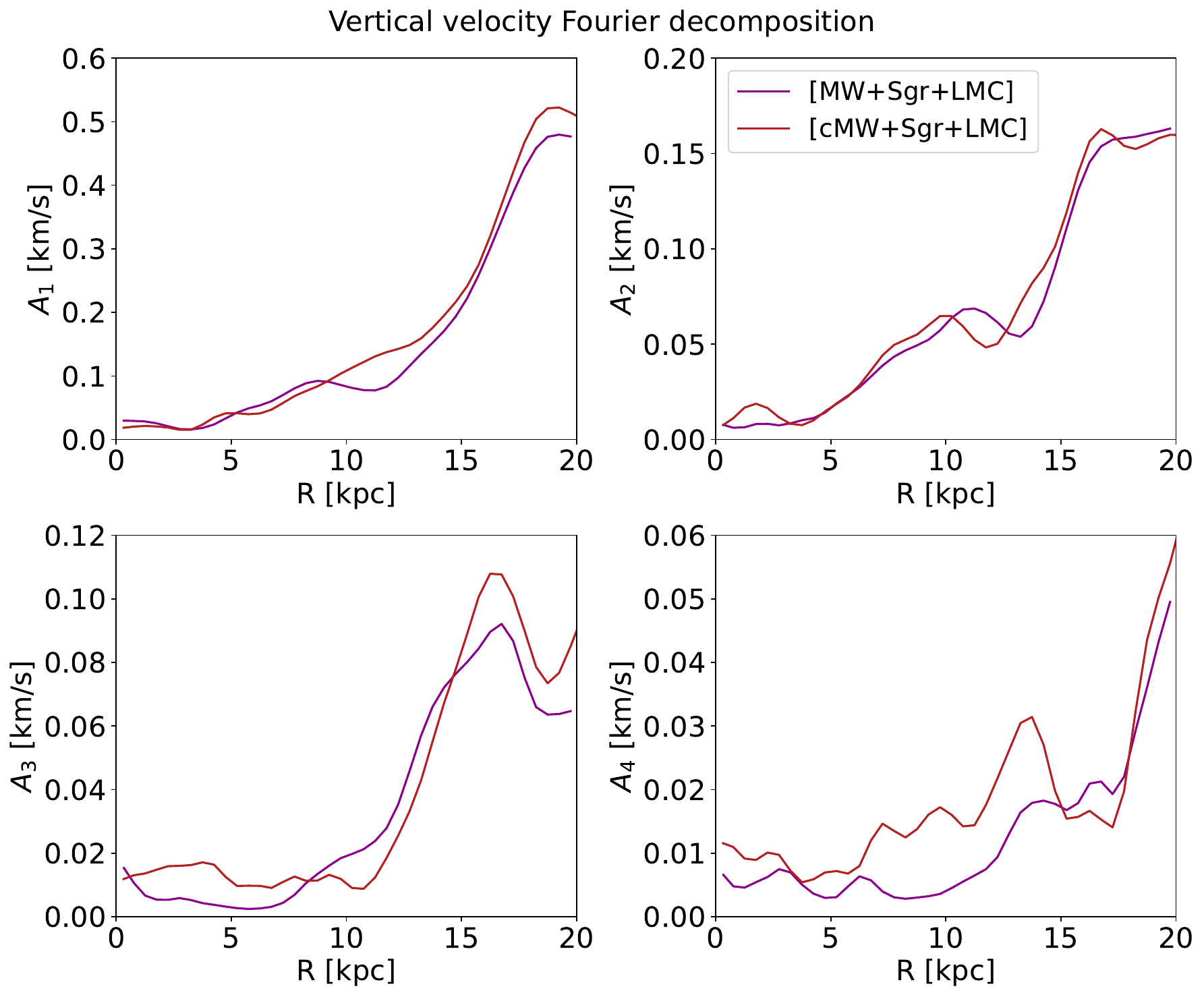}
    \caption{Same as Figure \ref{fig:pres-ff} but comparing Model [MW+Sgr+LMC] with the colder host disc in Model [cMW+Sgr+LMC].}
    \label{fig:cold_vert_pres}
  \end{figure}

Figure \ref{fig:cold_vert_pres} shows the Fourier decomposition for the `present day' mean vertical velocity fields for Model [cMW+Sgr+LMC] compared to Model [MW+Sgr+LMC]. In complete contrast to the radial velocity, the vertical velocity shows little difference between the warm and cold disc models (other than a slight overall increase in $A_4$), with the combined signatures of Sgr and the LMC discussed in Section \ref{sec:fouriervz} remaining clear.

\begin{figure}
    \centering
    \includegraphics[width=\linewidth]{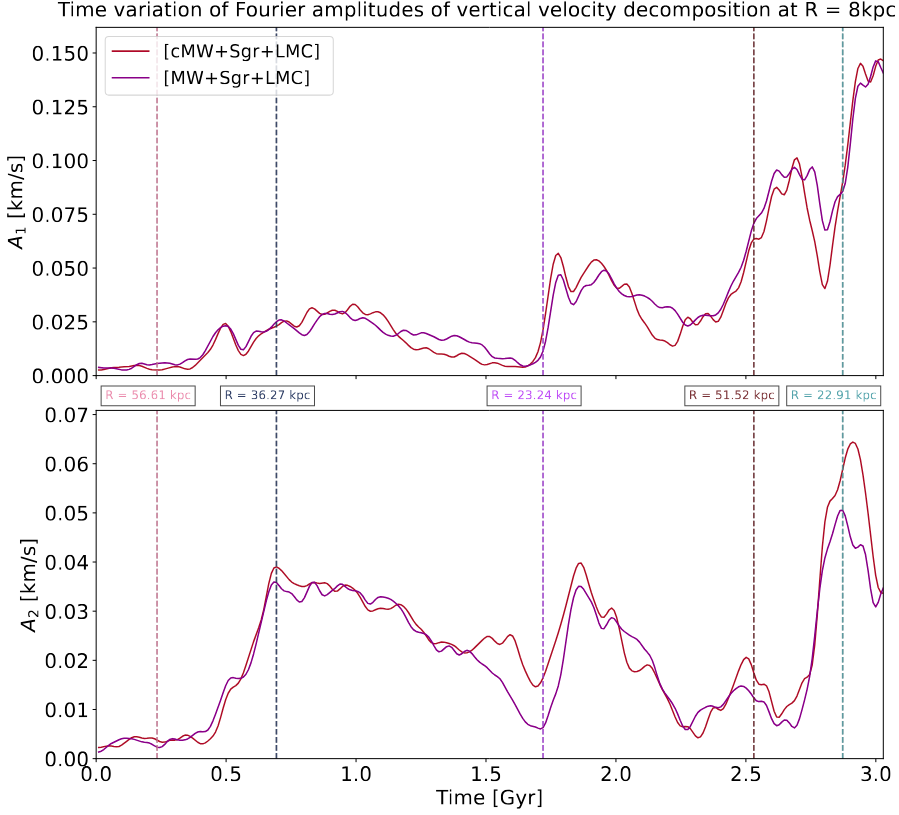}
    \caption{Same as Figure \ref{fig:vertime-ff} but comparing Model [MW+Sgr+LMC] with the colder host disc in Model [cMW+Sgr+LMC].}
    \label{fig:cold_vert_time}
\end{figure}

Figure \ref{fig:cold_vert_time} shows the Fourier decomposition for the mean vertical velocity fields at $R\sim8$ kpc over time for Model [cMW+Sgr+LMC] compared to Model [MW+Sgr+LMC]. Again, there is little difference in the vertical response to the Sgr impacts in $A_1$ or $A_2$ between the warm and cold disc models. As with the radial velocity, the $m=2$ mode induced by the second disc crossing mixes a little more slowly and retains a higher amplitude, $A_2$, prior to the third impact. Otherwise they are comparable.

Thus, the Fourier analysis matches the interpretation of the large scale $\bar{z}$ and $\bar{v}_z$ fields from Figure \ref{fig:colder}, suggesting the vertical signatures are more robust against the difference in disc velocity dispersion. Again, we remind the reader that as the disc does not buckle in either the warm or cold models and as such there are no significant features in the vertical velocity field without the influence of the satellites.

We are not suggesting that our model is a perfect reproduction of the vertical structure and dynamics of the disc, but that the features induced by both Sgr and the LMC should be visible across the disc of the Milky Way with next generation surveys such as SDSS-V Milky Way Mapper.

\section{Conclusions}

In this work we present a set of four high-resolution $N$-body numerical simulations of the interaction of the Milky Way, Sagittarius dwarf galaxy, and the LMC, following the orbital solutions of \cite{Vasiliev+21}. We present a systematic study of the way the satellites influence morphology and kinematics of the MW’s stellar disc, both individually and in combination, relative to an isolated disc scenario. Our results are a combination of confirming findings from previous lower resolution models, and new insights:

\textbf{Following other studies:}
\begin{itemize}
    \item[$-$] Each satellite has a distinct signature in the large scale density and velocity fields, where Sgr creates the ripples and corrugations, and the LMC creates a large scale warp, and both should be present in the Milky Way today \citep[see e.g.][and references within]{Laporte+18b_Sgr+LMC}.
    \item[$-$] Relating the warp induced by the LMC’s first in-fall to the warp observed in the outer disc is feasible but not conclusive. The orientation is roughly correct, but there remains a mismatch in both the amplitude and the angle of the line of nodes \citep[e.g.][]{Poggio+20,CabreraGadea+24,Jonsson+24}. This matches previous findings from \cite{Laporte+18a_LMC} despite a different model setup, suggesting that a dynamical influence in addition to that of the LMC is required to match the data or that the present models are too simplistic. For now, this remains an open question. 
\end{itemize}

\textbf{Our novel findings are:}
\begin{itemize}
    \item[$-$] The $\bar{v}_R$ corrugations in the stellar disc induced by a disc crossing of our Sgr model following the orbit from \cite{Vasiliev+21} naturally reproduce the correct location and phase of the radial velocity wave discovered in \cite{Eilers+20}. While correlation does not guarantee causation, our model implies that this wave may be a spiral feature driven by tidal interaction with the Sagittarius dwarf galaxy, and if so, may be used to further constrain the past orbit of Sgr.
    \item[$-$] Our simulations show the different signatures of Sgr and the LMC cleanly and in great detail owing to a combination of the resolution and the warm disc. The Fourier analysis reveals these overlapping modes are cumulative and separable, showing constructive or destructive interference in different regions.
    \item[$-$] Both the Sagittarius (Sgr) dwarf galaxy and the LMC have a relatively minor effect on the evolution of the galactic bar over the past 3 Gyr. The evolution of the bar remains largely undisturbed, with fluctuations in amplitude of no more than $\sim10$\% at the present day. This is in contrast to previous studies linking satellite impact with bar formation \citep[e.g.][]{Purcell+11}. This is examined further in our followup paper Petersen et al. (in prep).
    \item[$-$] While we show these signatures primarily in a warm stellar disc model, where they are clearest, the large scale patterns remain visible in a colder, more realistic stellar disc. This has implications for comparison with observations, where we would not expect a local region with secular fluctuations to well match the model \cite[e.g.][]{Vislosky+23}, but over large scales the imprint of both satellites should be apparent in next-generation surveys such as SDSS-V Milky Way Mapper \citep[and potentially as already seen in][]{Eilers+20}. 
\end{itemize}

Our findings showcase the potential of confronting simulations with observational data to unravel the intricate dynamics shaping the Milky Way’s disc. By combining our high-resolution simulations and future stellar surveys which are able to reveal large scale velocity fields of the Milky Way, we will be able to constrain the mass and orbital history of Sgr and the LMC, and disentangle their effect on the disc from secular structure evolution. 

\section*{acknowledgements}
IAS was supported by a Simons Foundation grant; `Supporting Development: Flatiron Research Fellows in mentoring and Local Undergraduates in Research'. KVJ was supported by NSF grant AST-1715582 and Simons Foundation grant 1018465. We thank Douglas Grion Filho for productive discussions around the Fourier decomposition methods and for providing the code. This research made use of \texttt{astropy}, a community-developed core Python package for Astronomy \citep{astropy-1,astropy-2}, the $N$-body tree code \texttt{Bonsai}\ \citep{Bonsai,Bonsai-242bil}, and the galactic dynamics Python packages \texttt{galpy} \citep{B15}, \texttt{Gala} \citep{gala} and \texttt{Agama} \citep{agama}. 

\bibliographystyle{aasjournal}
\bibliography{sample631}

\begin{thebibliography}{}
\expandafter\ifx\csname natexlab\endcsname\relax\def\natexlab#1{#1}\fi
\providecommand{\url}[1]{\href{#1}{#1}}
\providecommand{\dodoi}[1]{doi:~\href{http://doi.org/#1}{\nolinkurl{#1}}}
\providecommand{\doeprint}[1]{\href{http://ascl.net/#1}{\nolinkurl{http://ascl.net/#1}}}
\providecommand{\doarXiv}[1]{\href{https://arxiv.org/abs/#1}{\nolinkurl{https://arxiv.org/abs/#1}}}

\bibitem[{{Antoja} {et~al.}(2018){Antoja}, {Helmi}, {Romero-G{\'o}mez}, {Katz}, {Babusiaux}, {Drimmel}, {Evans}, {Figueras}, {Poggio}, {Reyl{\'e}}, {Robin}, {Seabroke}, \& {Soubiran}}]{Antoja+18}
{Antoja}, T., {Helmi}, A., {Romero-G{\'o}mez}, M., {et~al.} 2018, \nat, 561, 360, \dodoi{10.1038/s41586-018-0510-7}

\bibitem[{{Astropy Collaboration} {et~al.}(2013){Astropy Collaboration}, {Robitaille}, {Tollerud}, {Greenfield}, {Droettboom}, {Bray}, {Aldcroft}, {Davis}, {Ginsburg}, {Price-Whelan}, {Kerzendorf}, {Conley}, {Crighton}, {Barbary}, {Muna}, {Ferguson}, {Grollier}, {Parikh}, {Nair}, {Unther}, {Deil}, {Woillez}, {Conseil}, {Kramer}, {Turner}, {Singer}, {Fox}, {Weaver}, {Zabalza}, {Edwards}, {Azalee Bostroem}, {Burke}, {Casey}, {Crawford}, {Dencheva}, {Ely}, {Jenness}, {Labrie}, {Lim}, {Pierfederici}, {Pontzen}, {Ptak}, {Refsdal}, {Servillat}, \& {Streicher}}]{astropy-1}
{Astropy Collaboration}, {Robitaille}, T.~P., {Tollerud}, E.~J., {et~al.} 2013, \aap, 558, A33, \dodoi{10.1051/0004-6361/201322068}

\bibitem[{{Astropy Collaboration} {et~al.}(2018){Astropy Collaboration}, {Price-Whelan}, {Sip{\H{o}}cz}, {G{\"u}nther}, {Lim}, {Crawford}, {Conseil}, {Shupe}, {Craig}, {Dencheva}, {Ginsburg}, {Vand erPlas}, {Bradley}, {P{\'e}rez-Su{\'a}rez}, {de Val-Borro}, {Aldcroft}, {Cruz}, {Robitaille}, {Tollerud}, {Ardelean}, {Babej}, {Bach}, {Bachetti}, {Bakanov}, {Bamford}, {Barentsen}, {Barmby}, {Baumbach}, {Berry}, {Biscani}, {Boquien}, {Bostroem}, {Bouma}, {Brammer}, {Bray}, {Breytenbach}, {Buddelmeijer}, {Burke}, {Calderone}, {Cano Rodr{\'\i}guez}, {Cara}, {Cardoso}, {Cheedella}, {Copin}, {Corrales}, {Crichton}, {D'Avella}, {Deil}, {Depagne}, {Dietrich}, {Donath}, {Droettboom}, {Earl}, {Erben}, {Fabbro}, {Ferreira}, {Finethy}, {Fox}, {Garrison}, {Gibbons}, {Goldstein}, {Gommers}, {Greco}, {Greenfield}, {Groener}, {Grollier}, {Hagen}, {Hirst}, {Homeier}, {Horton}, {Hosseinzadeh}, {Hu}, {Hunkeler}, {Ivezi{\'c}}, {Jain}, {Jenness}, {Kanarek}, {Kendrew}, {Kern}, {Kerzendorf}, {Khvalko}, {King}, {Kirkby}, {Kulkarni},
  {Kumar}, {Lee}, {Lenz}, {Littlefair}, {Ma}, {Macleod}, {Mastropietro}, {McCully}, {Montagnac}, {Morris}, {Mueller}, {Mumford}, {Muna}, {Murphy}, {Nelson}, {Nguyen}, {Ninan}, {N{\"o}the}, {Ogaz}, {Oh}, {Parejko}, {Parley}, {Pascual}, {Patil}, {Patil}, {Plunkett}, {Prochaska}, {Rastogi}, {Reddy Janga}, {Sabater}, {Sakurikar}, {Seifert}, {Sherbert}, {Sherwood-Taylor}, {Shih}, {Sick}, {Silbiger}, {Singanamalla}, {Singer}, {Sladen}, {Sooley}, {Sornarajah}, {Streicher}, {Teuben}, {Thomas}, {Tremblay}, {Turner}, {Terr{\'o}n}, {van Kerkwijk}, {de la Vega}, {Watkins}, {Weaver}, {Whitmore}, {Woillez}, {Zabalza}, \& {Astropy Contributors}}]{astropy-2}
{Astropy Collaboration}, {Price-Whelan}, A.~M., {Sip{\H{o}}cz}, B.~M., {et~al.} 2018, \aj, 156, 123, \dodoi{10.3847/1538-3881/aabc4f}

\bibitem[{{B{\'e}dorf} {et~al.}(2014){B{\'e}dorf}, {Gaburov}, {Fujii}, {Nitadori}, {Ishiyama}, \& {Portegies Zwart}}]{Bonsai-242bil}
{B{\'e}dorf}, J., {Gaburov}, E., {Fujii}, M.~S., {et~al.} 2014, in Proceedings of the International Conference for High Performance Computing, 54--65, \dodoi{10.1109/SC.2014.10}

\bibitem[{{B{\'e}dorf} {et~al.}(2012){B{\'e}dorf}, {Gaburov}, \& {Portegies Zwart}}]{Bonsai}
{B{\'e}dorf}, J., {Gaburov}, E., \& {Portegies Zwart}, S. 2012, Journal of Computational Physics, 231, 2825, \dodoi{10.1016/j.jcp.2011.12.024}

\bibitem[{{Bennett} {et~al.}(2022){Bennett}, {Bovy}, \& {Hunt}}]{BBH21}
{Bennett}, M., {Bovy}, J., \& {Hunt}, J. A.~S. 2022, \apj, 927, 131, \dodoi{10.3847/1538-4357/ac5021}

\bibitem[{{Besla} {et~al.}(2007){Besla}, {Kallivayalil}, {Hernquist}, {Robertson}, {Cox}, {van der Marel}, \& {Alcock}}]{Besla+07}
{Besla}, G., {Kallivayalil}, N., {Hernquist}, L., {et~al.} 2007, \apj, 668, 949, \dodoi{10.1086/521385}

\bibitem[{{Besla} {et~al.}(2010){Besla}, {Kallivayalil}, {Hernquist}, {van der Marel}, {Cox}, \& {Kere{\v{s}}}}]{Besla+10}
---. 2010, \apjl, 721, L97, \dodoi{10.1088/2041-8205/721/2/L97}

\bibitem[{{Bland-Hawthorn} \& {Gerhard}(2016)}]{Bland-Hawthord+Gerhard16}
{Bland-Hawthorn}, J., \& {Gerhard}, O. 2016, \araa, 54, 529, \dodoi{10.1146/annurev-astro-081915-023441}

\bibitem[{{Blitz} \& {Spergel}(1991)}]{Blitz+Spergel91}
{Blitz}, L., \& {Spergel}, D.~N. 1991, \apj, 379, 631, \dodoi{10.1086/170535}

\bibitem[{{Bovy}(2015)}]{B15}
{Bovy}, J. 2015, \apjs, 216, 29, \dodoi{10.1088/0067-0049/216/2/29}

\bibitem[{{Bovy} {et~al.}(2019){Bovy}, {Leung}, {Hunt}, {Mackereth}, {Garc{\'\i}a-Hern{\'a}ndez}, \& {Roman-Lopes}}]{BLHM+19}
{Bovy}, J., {Leung}, H.~W., {Hunt}, J. A.~S., {et~al.} 2019, \mnras, 490, 4740, \dodoi{10.1093/mnras/stz2891}

\bibitem[{{Cabrera-Gadea} {et~al.}(2024){Cabrera-Gadea}, {Mateu}, {Ramos}, {Romero-G{\'o}mez}, {Antoja}, \& {Aguilar}}]{CabreraGadea+24}
{Cabrera-Gadea}, M., {Mateu}, C., {Ramos}, P., {et~al.} 2024, \mnras, 528, 4409, \dodoi{10.1093/mnras/stae308}

\bibitem[{{Choi} {et~al.}(2022){Choi}, {Olsen}, {Besla}, {van der Marel}, {Zivick}, {Kallivayalil}, \& {Nidever}}]{Choi+22}
{Choi}, Y., {Olsen}, K. A.~G., {Besla}, G., {et~al.} 2022, \apj, 927, 153, \dodoi{10.3847/1538-4357/ac4e90}

\bibitem[{{Cunningham} {et~al.}(2024){Cunningham}, {Hunt}, {Price-Whelan}, {Johnston}, {Ness}, {Lu}, {Escala}, \& {Stelea}}]{Cunningham+23}
{Cunningham}, E.~C., {Hunt}, J. A.~S., {Price-Whelan}, A.~M., {et~al.} 2024, \apj, 963, 95, \dodoi{10.3847/1538-4357/ad187b}

\bibitem[{{Das} {et~al.}(2024){Das}, {Huang}, {Ciuc{\u{a}}}, \& {Fragkoudi}}]{Das+23}
{Das}, P., {Huang}, Y., {Ciuc{\u{a}}}, I., \& {Fragkoudi}, F. 2024, \mnras, 527, 4505, \dodoi{10.1093/mnras/stad3344}

\bibitem[{{D'Onghia} \& {Fox}(2016)}]{donghia2016ARA&A..54..363D}
{D'Onghia}, E., \& {Fox}, A.~J. 2016, \araa, 54, 363, \dodoi{10.1146/annurev-astro-081915-023251}

\bibitem[{{Eilers} {et~al.}(2020){Eilers}, {Hogg}, {Rix}, {Frankel}, {Hunt}, {Fouvry}, \& {Buck}}]{Eilers+20}
{Eilers}, A.-C., {Hogg}, D.~W., {Rix}, H.-W., {et~al.} 2020, \apj, 900, 186, \dodoi{10.3847/1538-4357/abac0b}

\bibitem[{{Eilers} {et~al.}(2022){Eilers}, {Hogg}, {Rix}, {Ness}, {Price-Whelan}, {M{\'e}sz{\'a}ros}, \& {Nitschelm}}]{Eilers+22}
---. 2022, \apj, 928, 23, \dodoi{10.3847/1538-4357/ac54ad}

\bibitem[{{Erkal} {et~al.}(2019){Erkal}, {Belokurov}, {Laporte}, {Koposov}, {Li}, {Grillmair}, {Kallivayalil}, {Price-Whelan}, {Evans}, {Hawkins}, {Hendel}, {Mateu}, {Navarro}, {del Pino}, {Slater}, {Sohn}, \& {Orphan Aspen Treasury Collaboration}}]{Erkal+19}
{Erkal}, D., {Belokurov}, V., {Laporte}, C.~F.~P., {et~al.} 2019, \mnras, 487, 2685, \dodoi{10.1093/mnras/stz1371}

\bibitem[{{Foote} {et~al.}(2023){Foote}, {Besla}, {Mocz}, {Garavito-Camargo}, {Lancaster}, {Sparre}, {Cunningham}, {Vogelsberger}, {G{\'o}mez}, \& {Laporte}}]{Foote+23}
{Foote}, H.~R., {Besla}, G., {Mocz}, P., {et~al.} 2023, \apj, 954, 163, \dodoi{10.3847/1538-4357/ace533}

\bibitem[{{Frinchaboy} {et~al.}(2012){Frinchaboy}, {Majewski}, {Mu{\~n}oz}, {Law}, {{\L}okas}, {Kunkel}, {Patterson}, \& {Johnston}}]{Frinchaboy+12}
{Frinchaboy}, P.~M., {Majewski}, S.~R., {Mu{\~n}oz}, R.~R., {et~al.} 2012, \apj, 756, 74, \dodoi{10.1088/0004-637X/756/1/74}

\bibitem[{{Gaia Collaboration} {et~al.}(2016){Gaia Collaboration}, {Prusti}, {de Bruijne}, {Brown}, {Vallenari}, {Babusiaux}, {Bailer-Jones}, {Bastian}, {Biermann}, {Evans}, \& et~al.}]{GaiaMission}
{Gaia Collaboration}, {Prusti}, T., {de Bruijne}, J.~H.~J., {et~al.} 2016, \aap, 595, A1, \dodoi{10.1051/0004-6361/201629272}

\bibitem[{{Gaia Collaboration} {et~al.}(2018){Gaia Collaboration}, {Helmi}, {van Leeuwen}, {McMillan}, {Massari}, {Antoja}, {Robin}, {Lindegren}, {Bastian}, {Arenou}, {Babusiaux}, {Biermann}, {Breddels}, {Hobbs}, {Jordi}, {Pancino}, {Reyl{\'e}}, {Veljanoski}, {Brown}, {Vallenari}, {Prusti}, {de Bruijne}, {Bailer-Jones}, {Evans}, {Eyer}, {Jansen}, {Klioner}, {Lammers}, {Luri}, {Mignard}, {Panem}, {Pourbaix}, {Randich}, {Sartoretti}, {Siddiqui}, {Soubiran}, {Walton}, {Cropper}, {Drimmel}, {Katz}, {Lattanzi}, {Bakker}, {Cacciari}, {Casta{\~n}eda}, {Chaoul}, {Cheek}, {De Angeli}, {Fabricius}, {Guerra}, {Holl}, {Masana}, {Messineo}, {Mowlavi}, {Nienartowicz}, {Panuzzo}, {Portell}, {Riello}, {Seabroke}, {Tanga}, {Th{\'e}venin}, {Gracia-Abril}, {Comoretto}, {Garcia-Reinaldos}, {Teyssier}, {Altmann}, {Andrae}, {Audard}, {Bellas-Velidis}, {Benson}, {Berthier}, {Blomme}, {Burgess}, {Busso}, {Carry}, {Cellino}, {Clementini}, {Clotet}, {Creevey}, {Davidson}, {De Ridder}, {Delchambre}, {Dell'Oro}, {Ducourant},
  {Fern{\'a}ndez-Hern{\'a}ndez}, {Fouesneau}, {Fr{\'e}mat}, {Galluccio}, {Garc{\'\i}a-Torres}, {Gonz{\'a}lez-N{\'u}{\~n}ez}, {Gonz{\'a}lez-Vidal}, {Gosset}, {Guy}, {Halbwachs}, {Hambly}, {Harrison}, {Hern{\'a}ndez}, {Hestroffer}, {Hodgkin}, {Hutton}, {Jasniewicz}, {Jean-Antoine-Piccolo}, {Jordan}, {Korn}, {Krone-Martins}, {Lanzafame}, {Lebzelter}, {L{\"o}ffler}, {Manteiga}, {Marrese}, {Mart{\'\i}n-Fleitas}, {Moitinho}, {Mora}, {Muinonen}, {Osinde}, {Pauwels}, {Petit}, {Recio-Blanco}, {Richards}, {Rimoldini}, {Sarro}, {Siopis}, {Smith}, {Sozzetti}, {S{\"u}veges}, {Torra}, {van Reeven}, {Abbas}, {Abreu Aramburu}, {Accart}, {Aerts}, {Altavilla}, {{\'A}lvarez}, {Alvarez}, {Alves}, {Anderson}, {Andrei}, {Anglada Varela}, {Antiche}, {Arcay}, {Astraatmadja}, {Bach}, {Baker}, {Balaguer-N{\'u}{\~n}ez}, {Balm}, {Barache}, {Barata}, {Barbato}, {Barblan}, {Barklem}, {Barrado}, {Barros}, {Barstow}, {Bartholom{\'e} Mu{\~n}oz}, {Bassilana}, {Becciani}, {Bellazzini}, {Berihuete}, {Bertone}, {Bianchi}, {Bienaym{\'e}},
  {Blanco-Cuaresma}, {Boch}, {Boeche}, {Bombrun}, {Borrachero}, {Bossini}, {Bouquillon}, {Bourda}, {Bragaglia}, {Bramante}, {Bressan}, {Brouillet}, {Br{\"u}semeister}, {Brugaletta}, {Bucciarelli}, {Burlacu}, {Busonero}, {Butkevich}, {Buzzi}, {Caffau}, {Cancelliere}, {Cannizzaro}, {Cantat-Gaudin}, {Carballo}, {Carlucci}, {Carrasco}, {Casamiquela}, {Castellani}, {Castro-Ginard}, {Charlot}, {Chemin}, {Chiavassa}, {Cocozza}, {Costigan}, {Cowell}, {Crifo}, {Crosta}, {Crowley}, {Cuypers}, {Dafonte}, {Damerdji}, {Dapergolas}, {David}, {David}, {de Laverny}, {De Luise}, {De March}, {de Martino}, {de Souza}, {de Torres}, {Debosscher}, {del Pozo}, {Delbo}, {Delgado}, {Delgado}, {Di Matteo}, {Diakite}, {Diener}, {Distefano}, {Dolding}, {Drazinos}, {Dur{\'a}n}, {Edvardsson}, {Enke}, {Eriksson}, {Esquej}, {Eynard Bontemps}, {Fabre}, {Fabrizio}, {Faigler}, {Falc{\~a}o}, {Farr{\`a}s Casas}, {Federici}, {Fedorets}, {Fernique}, {Figueras}, {Filippi}, {Findeisen}, {Fonti}, {Fraile}, {Fraser}, {Fr{\'e}zouls}, {Gai}, {Galleti},
  {Garabato}, {Garc{\'\i}a-Sedano}, {Garofalo}, {Garralda}, {Gavel}, {Gavras}, {Gerssen}, {Geyer}, {Giacobbe}, {Gilmore}, {Girona}, {Giuffrida}, {Glass}, {Gomes}, {Granvik}, {Gueguen}, {Guerrier}, {Guiraud}, {Guti{\'e}rrez-S{\'a}nchez}, {Hofmann}, {Holland}, {Huckle}, {Hypki}, {Icardi}, {Jan{\ss}en}, {Jevardat de Fombelle}, {Jonker}, {Juh{\'a}sz}, {Julbe}, {Karampelas}, {Kewley}, {Klar}, {Kochoska}, {Kohley}, {Kolenberg}, {Kontizas}, {Kontizas}, {Koposov}, {Kordopatis}, {Kostrzewa-Rutkowska}, {Koubsky}, {Lambert}, {Lanza}, {Lasne}, {Lavigne}, {Le Fustec}, {Le Poncin-Lafitte}, {Lebreton}, {Leccia}, {Leclerc}, {Lecoeur-Taibi}, {Lenhardt}, {Leroux}, {Liao}, {Licata}, {Lindstr{\o}m}, {Lister}, {Livanou}, {Lobel}, {L{\'o}pez}, {Managau}, {Mann}, {Mantelet}, {Marchal}, {Marchant}, {Marconi}, {Marinoni}, {Marschalk{\'o}}, {Marshall}, {Martino}, {Marton}, {Mary}, {Matijevi{\v{c}}}, {Mazeh}, {Messina}, {Michalik}, {Millar}, {Molina}, {Molinaro}, {Moln{\'a}r}, {Montegriffo}, {Mor}, {Morbidelli}, {Morel}, {Morris},
  {Mulone}, {Muraveva}, {Musella}, {Nelemans}, {Nicastro}, {Noval}, {O'Mullane}, {Ord{\'e}novic}, {Ord{\'o}{\~n}ez-Blanco}, {Osborne}, {Pagani}, {Pagano}, {Pailler}, {Palacin}, {Palaversa}, {Panahi}, {Pawlak}, {Piersimoni}, {Pineau}, {Plachy}, {Plum}, {Poggio}, {Poujoulet}, {Pr{\v{s}}a}, {Pulone}, {Racero}, {Ragaini}, {Rambaux}, {Ramos-Lerate}, {Regibo}, {Riclet}, {Ripepi}, {Riva}, {Rivard}, {Rixon}, {Roegiers}, {Roelens}, {Romero-G{\'o}mez}, {Rowell}, {Royer}, {Ruiz-Dern}, {Sadowski}, {Sagrist{\`a} Sell{\'e}s}, {Sahlmann}, {Salgado}, {Salguero}, {Sanna}, {Santana-Ros}, {Sarasso}, {Savietto}, {Schultheis}, {Sciacca}, {Segol}, {Segovia}, {S{\'e}gransan}, {Shih}, {Siltala}, {Silva}, {Smart}, {Smith}, {Solano}, {Solitro}, {Sordo}, {Soria Nieto}, {Souchay}, {Spagna}, {Spoto}, {Stampa}, {Steele}, {Steidelm{\"u}ller}, {Stephenson}, {Stoev}, {Suess}, {Surdej}, {Szabados}, {Szegedi-Elek}, {Tapiador}, {Taris}, {Tauran}, {Taylor}, {Teixeira}, {Terrett}, {Teyssandier}, {Thuillot}, {Titarenko}, {Torra Clotet}, {Turon},
  {Ulla}, {Utrilla}, {Uzzi}, {Vaillant}, {Valentini}, {Valette}, {van Elteren}, {Van Hemelryck}, {van Leeuwen}, {Vaschetto}, {Vecchiato}, {Viala}, {Vicente}, {Vogt}, {von Essen}, {Voss}, {Votruba}, {Voutsinas}, {Walmsley}, {Weiler}, {Wertz}, {Wevems}, {Wyrzykowski}, {Yoldas}, {{\v{Z}}erjal}, {Ziaeepour}, {Zorec}, {Zschocke}, {Zucker}, {Zurbach}, \& {Zwitter}}]{Gaia_glob_dwarf+18}
{Gaia Collaboration}, {Helmi}, A., {van Leeuwen}, F., {et~al.} 2018, \aap, 616, A12, \dodoi{10.1051/0004-6361/201832698}

\bibitem[{{Gaia Collaboration} {et~al.}(2023{\natexlab{a}}){Gaia Collaboration}, {Drimmel}, {Romero-G{\'o}mez}, {Chemin}, {Ramos}, {Poggio}, {Ripepi}, {Andrae}, {Blomme}, {Cantat-Gaudin}, {Castro-Ginard}, {Clementini}, {Figueras}, {Fouesneau}, {Fr{\'e}mat}, {Jardine}, {Khanna}, {Lobel}, {Marshall}, {Muraveva}, {Brown}, {Vallenari}, {Prusti}, {de Bruijne}, {Arenou}, {Babusiaux}, {Biermann}, {Creevey}, {Ducourant}, {Evans}, {Eyer}, {Guerra}, {Hutton}, {Jordi}, {Klioner}, {Lammers}, {Lindegren}, {Luri}, {Mignard}, {Panem}, {Pourbaix}, {Randich}, {Sartoretti}, {Soubiran}, {Tanga}, {Walton}, {Bailer-Jones}, {Bastian}, {Jansen}, {Katz}, {Lattanzi}, {van Leeuwen}, {Bakker}, {Cacciari}, {Casta{\~n}eda}, {De Angeli}, {Fabricius}, {Galluccio}, {Guerrier}, {Heiter}, {Masana}, {Messineo}, {Mowlavi}, {Nicolas}, {Nienartowicz}, {Pailler}, {Panuzzo}, {Riclet}, {Roux}, {Seabroke}, {Sordo}, {Th{\'e}venin}, {Gracia-Abril}, {Portell}, {Teyssier}, {Altmann}, {Audard}, {Bellas-Velidis}, {Benson}, {Berthier}, {Burgess},
  {Busonero}, {Busso}, {C{\'a}novas}, {Carry}, {Cellino}, {Cheek}, {Damerdji}, {Davidson}, {de Teodoro}, {Nu{\~n}ez Campos}, {Delchambre}, {Dell'Oro}, {Esquej}, {Fern{\'a}ndez-Hern{\'a}ndez}, {Fraile}, {Garabato}, {Garc{\'\i}a-Lario}, {Gosset}, {Haigron}, {Halbwachs}, {Hambly}, {Harrison}, {Hern{\'a}ndez}, {Hestroffer}, {Hodgkin}, {Holl}, {Jan{\ss}en}, {Jevardat de Fombelle}, {Jordan}, {Krone-Martins}, {Lanzafame}, {L{\"o}ffler}, {Marchal}, {Marrese}, {Moitinho}, {Muinonen}, {Osborne}, {Pancino}, {Pauwels}, {Recio-Blanco}, {Reyl{\'e}}, {Riello}, {Rimoldini}, {Roegiers}, {Rybizki}, {Sarro}, {Siopis}, {Smith}, {Sozzetti}, {Utrilla}, {van Leeuwen}, {Abbas}, {{\'A}brah{\'a}m}, {Abreu Aramburu}, {Aerts}, {Aguado}, {Ajaj}, {Aldea-Montero}, {Altavilla}, {{\'A}lvarez}, {Alves}, {Anders}, {Anderson}, {Anglada Varela}, {Antoja}, {Baines}, {Baker}, {Balaguer-N{\'u}{\~n}ez}, {Balbinot}, {Balog}, {Barache}, {Barbato}, {Barros}, {Barstow}, {Bartolom{\'e}}, {Bassilana}, {Bauchet}, {Becciani}, {Bellazzini}, {Berihuete},
  {Bernet}, {Bertone}, {Bianchi}, {Binnenfeld}, {Blanco-Cuaresma}, {Boch}, {Bombrun}, {Bossini}, {Bouquillon}, {Bragaglia}, {Bramante}, {Breedt}, {Bressan}, {Brouillet}, {Brugaletta}, {Bucciarelli}, {Burlacu}, {Butkevich}, {Buzzi}, {Caffau}, {Cancelliere}, {Carballo}, {Carlucci}, {Carnerero}, {Carrasco}, {Casamiquela}, {Castellani}, {Chaoul}, {Charlot}, {Chiaramida}, {Chiavassa}, {Chornay}, {Comoretto}, {Contursi}, {Cooper}, {Cornez}, {Cowell}, {Crifo}, {Cropper}, {Crosta}, {Crowley}, {Dafonte}, {Dapergolas}, {David}, {de Laverny}, {De Luise}, {De March}, {De Ridder}, {de Souza}, {de Torres}, {del Peloso}, {del Pozo}, {Delbo}, {Delgado}, {Delisle}, {Demouchy}, {Dharmawardena}, {Di Matteo}, {Diakite}, {Diener}, {Distefano}, {Dolding}, {Enke}, {Fabre}, {Fabrizio}, {Faigler}, {Fedorets}, {Fernique}, {Fournier}, {Fouron}, {Fragkoudi}, {Gai}, {Garcia-Gutierrez}, {Garcia-Reinaldos}, {Garc{\'\i}a-Torres}, {Garofalo}, {Gavel}, {Gavras}, {Gerlach}, {Geyer}, {Giacobbe}, {Gilmore}, {Girona}, {Giuffrida}, {Gomel},
  {Gomez}, {Gonz{\'a}lez-N{\'u}{\~n}ez}, {Gonz{\'a}lez-Santamar{\'\i}a}, {Gonz{\'a}lez-Vidal}, {Granvik}, {Guillout}, {Guiraud}, {Guti{\'e}rrez-S{\'a}nchez}, {Guy}, {Hatzidimitriou}, {Hauser}, {Haywood}, {Helmer}, {Helmi}, {Sarmiento}, {Hidalgo}, {H{\l}adczuk}, {Hobbs}, {Holland}, {Huckle}, {Jasniewicz}, {Jean-Antoine Piccolo}, {Jim{\'e}nez-Arranz}, {Juaristi Campillo}, {Julbe}, {Karbevska}, {Kervella}, {Kordopatis}, {Korn}, {K{\'o}sp{\'a}l}, {Kostrzewa-Rutkowska}, {Kruszy{\'n}ska}, {Kun}, {Laizeau}, {Lambert}, {Lanza}, {Lasne}, {Le Campion}, {Lebreton}, {Lebzelter}, {Leccia}, {Leclerc}, {Lecoeur-Taibi}, {Liao}, {Licata}, {Lindstr{\o}m}, {Lister}, {Livanou}, {Lorca}, {Loup}, {Madrero Pardo}, {Magdaleno Romeo}, {Managau}, {Mann}, {Manteiga}, {Marchant}, {Marconi}, {Marcos}, {Marcos Santos}, {Mar{\'\i}n Pina}, {Marinoni}, {Marocco}, {Martin Polo}, {Mart{\'\i}n-Fleitas}, {Marton}, {Mary}, {Masip}, {Massari}, {Mastrobuono-Battisti}, {Mazeh}, {McMillan}, {Messina}, {Michalik}, {Millar}, {Mints}, {Molina},
  {Molinaro}, {Moln{\'a}r}, {Monari}, {Mongui{\'o}}, {Montegriffo}, {Montero}, {Mor}, {Mora}, {Morbidelli}, {Morel}, {Morris}, {Murphy}, {Musella}, {Nagy}, {Noval}, {Oca{\~n}a}, {Ogden}, {Ordenovic}, {Osinde}, {Pagani}, {Pagano}, {Palaversa}, {Palicio}, {Pallas-Quintela}, {Panahi}, {Payne-Wardenaar}, {Pe{\~n}alosa Esteller}, {Penttil{\"a}}, {Pichon}, {Piersimoni}, {Pineau}, {Plachy}, {Plum}, {Pr{\v{s}}a}, {Pulone}, {Racero}, {Ragaini}, {Rainer}, {Raiteri}, {Ramos-Lerate}, {Re Fiorentin}, {Regibo}, {Richards}, {Rios Diaz}, {Riva}, {Rix}, {Rixon}, {Robichon}, {Robin}, {Robin}, {Roelens}, {Rogues}, {Rohrbasser}, {Rowell}, {Royer}, {Ruz Mieres}, {Rybicki}, {Sadowski}, {S{\'a}ez N{\'u}{\~n}ez}, {Sagrist{\`a} Sell{\'e}s}, {Sahlmann}, {Salguero}, {Samaras}, {Sanchez Gimenez}, {Sanna}, {Santove{\~n}a}, {Sarasso}, {Schultheis}, {Sciacca}, {Segol}, {Segovia}, {S{\'e}gransan}, {Semeux}, {Shahaf}, {Siddiqui}, {Siebert}, {Siltala}, {Silvelo}, {Slezak}, {Slezak}, {Smart}, {Snaith}, {Solano}, {Solitro}, {Souami}, {Souchay},
  {Spagna}, {Spina}, {Spoto}, {Steele}, {Steidelm{\"u}ller}, {Stephenson}, {S{\"u}veges}, {Surdej}, {Szabados}, {Szegedi-Elek}, {Taris}, {Taylor}, {Teixeira}, {Tolomei}, {Tonello}, {Torra}, {Torra}, {Torralba Elipe}, {Trabucchi}, {Tsounis}, {Turon}, {Ulla}, {Unger}, {Vaillant}, {van Dillen}, {van Reeven}, {Vanel}, {Vecchiato}, {Viala}, {Vicente}, {Voutsinas}, {Weiler}, {Wevers}, {Wyrzykowski}, {Yoldas}, {Yvard}, {Zhao}, {Zorec}, {Zucker}, \& {Zwitter}}]{GaiaCollab+23_mapasym}
{Gaia Collaboration}, {Drimmel}, R., {Romero-G{\'o}mez}, M., {et~al.} 2023{\natexlab{a}}, \aap, 674, A37, \dodoi{10.1051/0004-6361/202243797}

\bibitem[{{Gaia Collaboration} {et~al.}(2023{\natexlab{b}}){Gaia Collaboration}, {Recio-Blanco}, {Kordopatis}, {de Laverny}, {Palicio}, {Spagna}, {Spina}, {Katz}, {Re Fiorentin}, {Poggio}, {McMillan}, {Vallenari}, {Lattanzi}, {Seabroke}, {Casamiquela}, {Bragaglia}, {Antoja}, {Bailer-Jones}, {Schultheis}, {Andrae}, {Fouesneau}, {Cropper}, {Cantat-Gaudin}, {Bijaoui}, {Heiter}, {Brown}, {Prusti}, {de Bruijne}, {Arenou}, {Babusiaux}, {Biermann}, {Creevey}, {Ducourant}, {Evans}, {Eyer}, {Guerra}, {Hutton}, {Jordi}, {Klioner}, {Lammers}, {Lindegren}, {Luri}, {Mignard}, {Panem}, {Pourbaix}, {Randich}, {Sartoretti}, {Soubiran}, {Tanga}, {Walton}, {Bastian}, {Drimmel}, {Jansen}, {van Leeuwen}, {Bakker}, {Cacciari}, {Casta{\~n}eda}, {De Angeli}, {Fabricius}, {Fr{\'e}mat}, {Galluccio}, {Guerrier}, {Masana}, {Messineo}, {Mowlavi}, {Nicolas}, {Nienartowicz}, {Pailler}, {Panuzzo}, {Riclet}, {Roux}, {Sordo}, {Th{\'e}venin}, {Gracia-Abril}, {Portell}, {Teyssier}, {Altmann}, {Audard}, {Bellas-Velidis}, {Benson}, {Berthier},
  {Blomme}, {Burgess}, {Busonero}, {Busso}, {C{\'a}novas}, {Carry}, {Cellino}, {Cheek}, {Clementini}, {Damerdji}, {Davidson}, {de Teodoro}, {Nu{\~n}ez Campos}, {Delchambre}, {Dell'Oro}, {Esquej}, {Fern{\'a}ndez-Hern{\'a}ndez}, {Fraile}, {Garabato}, {Garc{\'\i}a-Lario}, {Gosset}, {Haigron}, {Halbwachs}, {Hambly}, {Harrison}, {Hern{\'a}ndez}, {Hestroffer}, {Hodgkin}, {Holl}, {Jan{\ss}en}, {Jevardat de Fombelle}, {Jordan}, {Krone-Martins}, {Lanzafame}, {L{\"o}ffler}, {Marchal}, {Marrese}, {Moitinho}, {Muinonen}, {Osborne}, {Pancino}, {Pauwels}, {Reyl{\'e}}, {Riello}, {Rimoldini}, {Roegiers}, {Rybizki}, {Sarro}, {Siopis}, {Smith}, {Sozzetti}, {Utrilla}, {van Leeuwen}, {Abbas}, {{\'A}brah{\'a}m}, {Abreu Aramburu}, {Aerts}, {Aguado}, {Ajaj}, {Aldea-Montero}, {Altavilla}, {{\'A}lvarez}, {Alves}, {Anders}, {Anderson}, {Anglada Varela}, {Baines}, {Baker}, {Balaguer-N{\'u}{\~n}ez}, {Balbinot}, {Balog}, {Barache}, {Barbato}, {Barros}, {Barstow}, {Bartolom{\'e}}, {Bassilana}, {Bauchet}, {Becciani}, {Bellazzini},
  {Berihuete}, {Bernet}, {Bertone}, {Bianchi}, {Binnenfeld}, {Blanco-Cuaresma}, {Boch}, {Bombrun}, {Bossini}, {Bouquillon}, {Bramante}, {Breedt}, {Bressan}, {Brouillet}, {Brugaletta}, {Bucciarelli}, {Burlacu}, {Butkevich}, {Buzzi}, {Caffau}, {Cancelliere}, {Carballo}, {Carlucci}, {Carnerero}, {Carrasco}, {Castellani}, {Castro-Ginard}, {Chaoul}, {Charlot}, {Chemin}, {Chiaramida}, {Chiavassa}, {Chornay}, {Comoretto}, {Contursi}, {Cooper}, {Cornez}, {Cowell}, {Crifo}, {Crosta}, {Crowley}, {Dafonte}, {Dapergolas}, {David}, {De Luise}, {De March}, {De Ridder}, {de Souza}, {de Torres}, {del Peloso}, {del Pozo}, {Delbo}, {Delgado}, {Delisle}, {Demouchy}, {Dharmawardena}, {Di Matteo}, {Diakite}, {Diener}, {Distefano}, {Dolding}, {Edvardsson}, {Enke}, {Fabre}, {Fabrizio}, {Faigler}, {Fedorets}, {Fernique}, {Figueras}, {Fournier}, {Fouron}, {Fragkoudi}, {Gai}, {Garcia-Gutierrez}, {Garcia-Reinaldos}, {Garc{\'\i}a-Torres}, {Garofalo}, {Gavel}, {Gavras}, {Gerlach}, {Geyer}, {Giacobbe}, {Gilmore}, {Girona}, {Giuffrida},
  {Gomel}, {Gomez}, {Gonz{\'a}lez-N{\'u}{\~n}ez}, {Gonz{\'a}lez-Santamar{\'\i}a}, {Gonz{\'a}lez-Vidal}, {Granvik}, {Guillout}, {Guiraud}, {Guti{\'e}rrez-S{\'a}nchez}, {Guy}, {Hatzidimitriou}, {Hauser}, {Haywood}, {Helmer}, {Helmi}, {Sarmiento}, {Hidalgo}, {H{\l}adczuk}, {Hobbs}, {Holland}, {Huckle}, {Jardine}, {Jasniewicz}, {Jean-Antoine Piccolo}, {Jim{\'e}nez-Arranz}, {Juaristi Campillo}, {Julbe}, {Karbevska}, {Kervella}, {Khanna}, {Korn}, {K{\'o}sp{\'a}l}, {Kostrzewa-Rutkowska}, {Kruszy{\'n}ska}, {Kun}, {Laizeau}, {Lambert}, {Lanza}, {Lasne}, {Le Campion}, {Lebreton}, {Lebzelter}, {Leccia}, {Leclerc}, {Lecoeur-Taibi}, {Liao}, {Licata}, {Lindstr{\o}m}, {Lister}, {Livanou}, {Lobel}, {Lorca}, {Loup}, {Madrero Pardo}, {Magdaleno Romeo}, {Managau}, {Mann}, {Manteiga}, {Marchant}, {Marconi}, {Marcos}, {Marcos Santos}, {Mar{\'\i}n Pina}, {Marinoni}, {Marocco}, {Marshall}, {Martin Polo}, {Mart{\'\i}n-Fleitas}, {Marton}, {Mary}, {Masip}, {Massari}, {Mastrobuono-Battisti}, {Mazeh}, {Messina}, {Michalik}, {Millar},
  {Mints}, {Molina}, {Molinaro}, {Moln{\'a}r}, {Monari}, {Mongui{\'o}}, {Montegriffo}, {Montero}, {Mor}, {Mora}, {Morbidelli}, {Morel}, {Morris}, {Muraveva}, {Murphy}, {Musella}, {Nagy}, {Noval}, {Oca{\~n}a}, {Ogden}, {Ordenovic}, {Osinde}, {Pagani}, {Pagano}, {Palaversa}, {Pallas-Quintela}, {Panahi}, {Payne-Wardenaar}, {Pe{\~n}alosa Esteller}, {Penttil{\"a}}, {Pichon}, {Piersimoni}, {Pineau}, {Plachy}, {Plum}, {Pr{\v{s}}a}, {Pulone}, {Racero}, {Ragaini}, {Rainer}, {Raiteri}, {Ramos}, {Ramos-Lerate}, {Regibo}, {Richards}, {Rios Diaz}, {Ripepi}, {Riva}, {Rix}, {Rixon}, {Robichon}, {Robin}, {Robin}, {Roelens}, {Rogues}, {Rohrbasser}, {Romero-G{\'o}mez}, {Rowell}, {Royer}, {Ruz Mieres}, {Rybicki}, {Sadowski}, {S{\'a}ez N{\'u}{\~n}ez}, {Sagrist{\`a} Sell{\'e}s}, {Sahlmann}, {Salguero}, {Samaras}, {Sanchez Gimenez}, {Sanna}, {Santove{\~n}a}, {Sarasso}, {Sciacca}, {Segol}, {Segovia}, {S{\'e}gransan}, {Semeux}, {Shahaf}, {Siddiqui}, {Siebert}, {Siltala}, {Silvelo}, {Slezak}, {Slezak}, {Smart}, {Snaith}, {Solano},
  {Solitro}, {Souami}, {Souchay}, {Spoto}, {Steele}, {Steidelm{\"u}ller}, {Stephenson}, {S{\"u}veges}, {Surdej}, {Szabados}, {Szegedi-Elek}, {Taris}, {Taylor}, {Teixeira}, {Tolomei}, {Tonello}, {Torra}, {Torra}, {Torralba Elipe}, {Trabucchi}, {Tsounis}, {Turon}, {Ulla}, {Unger}, {Vaillant}, {van Dillen}, {van Reeven}, {Vanel}, {Vecchiato}, {Viala}, {Vicente}, {Voutsinas}, {Weiler}, {Wevers}, {Wyrzykowski}, {Yoldas}, {Yvard}, {Zhao}, {Zorec}, {Zucker}, \& {Zwitter}}]{DR3_chemcart}
{Gaia Collaboration}, {Recio-Blanco}, A., {Kordopatis}, G., {et~al.} 2023{\natexlab{b}}, \aap, 674, A38, \dodoi{10.1051/0004-6361/202243511}

\bibitem[{{Garavito-Camargo} {et~al.}(2019){Garavito-Camargo}, {Besla}, {Laporte}, {Johnston}, {G{\'o}mez}, \& {Watkins}}]{Garavito-Camargo+19}
{Garavito-Camargo}, N., {Besla}, G., {Laporte}, C. F.~P., {et~al.} 2019, \apj, 884, 51, \dodoi{10.3847/1538-4357/ab32eb}

\bibitem[{{Garavito-Camargo} {et~al.}(2021){Garavito-Camargo}, {Patel}, {Besla}, {Price-Whelan}, {G{\'o}mez}, {Laporte}, \& {Johnston}}]{Garavito-Camargo+21}
{Garavito-Camargo}, N., {Patel}, E., {Besla}, G., {et~al.} 2021, \apj, 923, 140, \dodoi{10.3847/1538-4357/ac2c05}

\bibitem[{{Garavito-Camargo} {et~al.}(2023){Garavito-Camargo}, {Price-Whelan}, {Samuel}, {Cunningham}, {Patel}, {Wetzel}, {Johnston}, {Arora}, {Sanderson}, {Garrison}, \& {Horta}}]{Garavito-Camargo+23}
{Garavito-Camargo}, N., {Price-Whelan}, A.~M., {Samuel}, J., {et~al.} 2023, arXiv e-prints, arXiv:2311.11359, \dodoi{10.48550/arXiv.2311.11359}

\bibitem[{{Gardiner} \& {Noguchi}(1996)}]{Gardiner+96}
{Gardiner}, L.~T., \& {Noguchi}, M. 1996, \mnras, 278, 191, \dodoi{10.1093/mnras/278.1.191}

\bibitem[{{Gibbons} {et~al.}(2017){Gibbons}, {Belokurov}, \& {Evans}}]{Gibbons+17}
{Gibbons}, S.~L.~J., {Belokurov}, V., \& {Evans}, N.~W. 2017, \mnras, 464, 794, \dodoi{10.1093/mnras/stw2328}

\bibitem[{{G{\'o}mez} {et~al.}(2015){G{\'o}mez}, {Besla}, {Carpintero}, {Villalobos}, {O'Shea}, \& {Bell}}]{gomez2015ApJ...802..128G}
{G{\'o}mez}, F.~A., {Besla}, G., {Carpintero}, D.~D., {et~al.} 2015, \apj, 802, 128, \dodoi{10.1088/0004-637X/802/2/128}

\bibitem[{{G{\'o}mez} {et~al.}(2013){G{\'o}mez}, {Minchev}, {O'Shea}, {Beers}, {Bullock}, \& {Purcell}}]{gomez2013MNRAS.429..159G}
{G{\'o}mez}, F.~A., {Minchev}, I., {O'Shea}, B.~W., {et~al.} 2013, \mnras, 429, 159, \dodoi{10.1093/mnras/sts327}

\bibitem[{{G{\'o}mez} {et~al.}(2016){G{\'o}mez}, {White}, {Marinacci}, {Slater}, {Grand}, {Springel}, \& {Pakmor}}]{Gomez+16}
{G{\'o}mez}, F.~A., {White}, S. D.~M., {Marinacci}, F., {et~al.} 2016, \mnras, 456, 2779, \dodoi{10.1093/mnras/stv2786}

\bibitem[{{Grion Filho} {et~al.}(2020){Grion Filho}, {Johnston}, {Poggio}, {Laporte}, {Drimmel}, \& {D'Onghia}}]{GrionFilho+20}
{Grion Filho}, D., {Johnston}, K.~V., {Poggio}, E., {et~al.} 2020, arXiv e-prints, arXiv:2012.07778.
\newblock \doarXiv{2012.07778}

\bibitem[{{Han} {et~al.}(2023{\natexlab{a}}){Han}, {Conroy}, \& {Hernquist}}]{Han+23_tiltwarp}
{Han}, J.~J., {Conroy}, C., \& {Hernquist}, L. 2023{\natexlab{a}}, Nature Astronomy, \dodoi{10.1038/s41550-023-02076-9}

\bibitem[{{Han} {et~al.}(2023{\natexlab{b}}){Han}, {Semenov}, {Conroy}, \& {Hernquist}}]{Han+24}
{Han}, J.~J., {Semenov}, V., {Conroy}, C., \& {Hernquist}, L. 2023{\natexlab{b}}, \apjl, 957, L24, \dodoi{10.3847/2041-8213/ad0641}

\bibitem[{{Heller} \& {Rohlfs}(1994)}]{Heller+94}
{Heller}, P., \& {Rohlfs}, K. 1994, \aap, 291, 743

\bibitem[{{Hey} {et~al.}(2023){Hey}, {Huber}, {Shappee}, {Bland-Hawthorn}, {Tepper-Garc{\'\i}a}, {Sanderson}, {Chakrabarti}, {Saunders}, {Hunt}, {Bedding}, \& {Tonry}}]{Hey+23}
{Hey}, D.~R., {Huber}, D., {Shappee}, B.~J., {et~al.} 2023, \aj, 166, 249, \dodoi{10.3847/1538-3881/ad01bf}

\bibitem[{{Hrannar J{\'o}nsson} \& {McMillan}(2024)}]{Jonsson+24}
{Hrannar J{\'o}nsson}, V., \& {McMillan}, P.~J. 2024, arXiv e-prints, arXiv:2405.09624, \dodoi{10.48550/arXiv.2405.09624}

\bibitem[{{Hunt} {et~al.}(2021){Hunt}, {Stelea}, {Johnston}, {Gandhi}, {Laporte}, \& {B{\'e}dorf}}]{Hunt+21}
{Hunt}, J. A.~S., {Stelea}, I.~A., {Johnston}, K.~V., {et~al.} 2021, \mnras, 508, 1459, \dodoi{10.1093/mnras/stab2580}

\bibitem[{{Jim{\'e}nez-Arranz} {et~al.}(2024){Jim{\'e}nez-Arranz}, {Roca-F{\`a}brega}, {Romero-G{\'o}mez}, {Luri}, {Bernet}, {McMillan}, \& {Chemin}}]{JimenezArranz+24}
{Jim{\'e}nez-Arranz}, {\'O}., {Roca-F{\`a}brega}, S., {Romero-G{\'o}mez}, M., {et~al.} 2024, \aap, 688, A51, \dodoi{10.1051/0004-6361/202349058}

\bibitem[{{Kallivayalil} {et~al.}(2013){Kallivayalil}, {van der Marel}, {Besla}, {Anderson}, \& {Alcock}}]{Kallivayalil+13}
{Kallivayalil}, N., {van der Marel}, R.~P., {Besla}, G., {Anderson}, J., \& {Alcock}, C. 2013, \apj, 764, 161, \dodoi{10.1088/0004-637X/764/2/161}

\bibitem[{{Kawata} {et~al.}(2018){Kawata}, {Baba}, {Ciuc{\v a}}, {Cropper}, {Grand}, {Hunt}, \& {Seabroke}}]{KBCCGHS18}
{Kawata}, D., {Baba}, J., {Ciuc{\v a}}, I., {et~al.} 2018, \mnras, 479, L108, \dodoi{10.1093/mnrasl/sly107}

\bibitem[{{Khanna} {et~al.}(2019){Khanna}, {Sharma}, {Tepper-Garcia}, {Bland -Hawthorn}, {Hayden}, {Asplund}, {Buder}, {Chen}, {De Silva}, {Freeman}, {Kos}, {Lewis}, {Lin}, {Martell}, {Simpson}, {Nordlander}, {Stello}, {Ting}, {Zucker}, \& {Zwitter}}]{Khanna+19}
{Khanna}, S., {Sharma}, S., {Tepper-Garcia}, T., {et~al.} 2019, \mnras, 489, 4962, \dodoi{10.1093/mnras/stz2462}

\bibitem[{{King}(1962)}]{King62}
{King}, I. 1962, \aj, 67, 471, \dodoi{10.1086/108756}

\bibitem[{{Kollmeier} {et~al.}(2019){Kollmeier}, {Anderson}, {Blanc}, {Blanton}, {Covey}, {Crane}, {Drory}, {Frinchaboy}, {Froning}, {Johnson}, {Kneib}, {Kreckel}, {Merloni}, {Pellegrini}, {Pogge}, {Ramirez}, {Rix}, {Sayres}, {S{\'a}nchez-Gallego}, {Shen}, {Tkachenko}, {Trump}, {Tuttle}, {Weijmans}, {Zasowski}, {Barbuy}, {Beaton}, {Bergemann}, {Bochanski}, {Brandt}, {Casey}, {Cherinka}, {Eracleous}, {Fan}, {Garc{\'\i}a}, {Green}, {Hekker}, {Lane}, {Longa-Pe{\~n}a}, {Mathur}, {Meza}, {Minchev}, {Myers}, {Nidever}, {Nitschelm}, {O'Connell}, {Price-Whelan}, {Raddick}, {Rossi}, {Sankrit}, {Simon}, {Stutz}, {Ting}, {Trakhtenbrot}, {Weaver}, {Willmer}, \& {Weinberg}}]{Kollmeier+19}
{Kollmeier}, J., {Anderson}, S.~F., {Blanc}, G.~A., {et~al.} 2019, \baas, 51, 274

\bibitem[{{Laporte} {et~al.}(2018{\natexlab{a}}){Laporte}, {G{\'o}mez}, {Besla}, {Johnston}, \& {Garavito-Camargo}}]{Laporte+18a_LMC}
{Laporte}, C. F.~P., {G{\'o}mez}, F.~A., {Besla}, G., {Johnston}, K.~V., \& {Garavito-Camargo}, N. 2018{\natexlab{a}}, \mnras, 473, 1218, \dodoi{10.1093/mnras/stx2146}

\bibitem[{{Laporte} {et~al.}(2018{\natexlab{b}}){Laporte}, {Johnston}, {G{\'o}mez}, {Garavito-Camargo}, \& {Besla}}]{Laporte+18b_Sgr+LMC}
{Laporte}, C.~F.~P., {Johnston}, K.~V., {G{\'o}mez}, F.~A., {Garavito-Camargo}, N., \& {Besla}, G. 2018{\natexlab{b}}, \mnras, 481, 286, \dodoi{10.1093/mnras/sty1574}

\bibitem[{{Laporte} {et~al.}(2022){Laporte}, {Koposov}, \& {Belokurov}}]{Laporte+22_feathers}
{Laporte}, C. F.~P., {Koposov}, S.~E., \& {Belokurov}, V. 2022, \mnras, 510, L13, \dodoi{10.1093/mnrasl/slab109}

\bibitem[{{Laporte} {et~al.}(2019){Laporte}, {Minchev}, {Johnston}, \& {G{\'o}mez}}]{Laporte+19_Gaia}
{Laporte}, C.~F.~P., {Minchev}, I., {Johnston}, K.~V., \& {G{\'o}mez}, F.~A. 2019, \mnras, 485, 3134, \dodoi{10.1093/mnras/stz583}

\bibitem[{{Leung} {et~al.}(2023){Leung}, {Bovy}, {Mackereth}, {Hunt}, {Lane}, \& {Wilson}}]{Leung+23}
{Leung}, H.~W., {Bovy}, J., {Mackereth}, J.~T., {et~al.} 2023, \mnras, 519, 948, \dodoi{10.1093/mnras/stac3529}

\bibitem[{{Levine} {et~al.}(2006){Levine}, {Blitz}, \& {Heiles}}]{Levine+06}
{Levine}, E.~S., {Blitz}, L., \& {Heiles}, C. 2006, \apj, 643, 881, \dodoi{10.1086/503091}

\bibitem[{{Lilleengen} {et~al.}(2023){Lilleengen}, {Petersen}, {Erkal}, {Pe{\~n}arrubia}, {Koposov}, {Li}, {Cullinane}, {Ji}, {Kuehn}, {Lewis}, {Mackey}, {Pace}, {Shipp}, {Zucker}, {Bland-Hawthorn}, {Hilmi}, \& {S5 Collaboration}}]{Lilleengen+23}
{Lilleengen}, S., {Petersen}, M.~S., {Erkal}, D., {et~al.} 2023, \mnras, 518, 774, \dodoi{10.1093/mnras/stac3108}

\bibitem[{{Majewski} {et~al.}(2016){Majewski}, {APOGEE Team}, \& {APOGEE-2 Team}}]{MAPOGEE16}
{Majewski}, S.~R., {APOGEE Team}, \& {APOGEE-2 Team}. 2016, Astronomische Nachrichten, 337, 863, \dodoi{10.1002/asna.201612387}

\bibitem[{{Pe{\~n}arrubia} {et~al.}(2016){Pe{\~n}arrubia}, {G{\'o}mez}, {Besla}, {Erkal}, \& {Ma}}]{Penarubia2016}
{Pe{\~n}arrubia}, J., {G{\'o}mez}, F.~A., {Besla}, G., {Erkal}, D., \& {Ma}, Y.-Z. 2016, \mnras, 456, L54, \dodoi{10.1093/mnrasl/slv160}

\bibitem[{{Petersen} \& {Pe{\~n}arrubia}(2021)}]{Petersen+Penarrubia21}
{Petersen}, M.~S., \& {Pe{\~n}arrubia}, J. 2021, Nature Astronomy, 5, 251, \dodoi{10.1038/s41550-020-01254-3}

\bibitem[{{Poggio} {et~al.}(2020){Poggio}, {Drimmel}, {Andrae}, {Bailer-Jones}, {Fouesneau}, {Lattanzi}, {Smart}, \& {Spagna}}]{Poggio+20}
{Poggio}, E., {Drimmel}, R., {Andrae}, R., {et~al.} 2020, Nature Astronomy, 4, 590, \dodoi{10.1038/s41550-020-1017-3}

\bibitem[{{Poggio} {et~al.}(2021){Poggio}, {Laporte}, {Johnston}, {D'Onghia}, {Drimmel}, \& {Grion Filho}}]{Poggio+21}
{Poggio}, E., {Laporte}, C. F.~P., {Johnston}, K.~V., {et~al.} 2021, \mnras, 508, 541, \dodoi{10.1093/mnras/stab2245}

\bibitem[{{Price-Whelan}(2017)}]{gala}
{Price-Whelan}, A.~M. 2017, The Journal of Open Source Software, 2, 388, \dodoi{10.21105/joss.00388}

\bibitem[{{Price-Whelan} {et~al.}(2015){Price-Whelan}, {Johnston}, {Sheffield}, {Laporte}, \& {Sesar}}]{Price-Whelan+15}
{Price-Whelan}, A.~M., {Johnston}, K.~V., {Sheffield}, A.~A., {Laporte}, C. F.~P., \& {Sesar}, B. 2015, \mnras, 452, 676, \dodoi{10.1093/mnras/stv1324}

\bibitem[{{Purcell} {et~al.}(2011){Purcell}, {Bullock}, {Tollerud}, {Rocha}, \& {Chakrabarti}}]{Purcell+11}
{Purcell}, C.~W., {Bullock}, J.~S., {Tollerud}, E.~J., {Rocha}, M., \& {Chakrabarti}, S. 2011, \nat, 477, 301, \dodoi{10.1038/nature10417}

\bibitem[{{Ruiz-Lara} {et~al.}(2020){Ruiz-Lara}, {Gallart}, {Bernard}, \& {Cassisi}}]{RuizLara+20}
{Ruiz-Lara}, T., {Gallart}, C., {Bernard}, E.~J., \& {Cassisi}, S. 2020, Nature Astronomy, 4, 965, \dodoi{10.1038/s41550-020-1097-0}

\bibitem[{{van der Marel} {et~al.}(2002){van der Marel}, {Alves}, {Hardy}, \& {Suntzeff}}]{vanderMarel+02}
{van der Marel}, R.~P., {Alves}, D.~R., {Hardy}, E., \& {Suntzeff}, N.~B. 2002, \aj, 124, 2639, \dodoi{10.1086/343775}

\bibitem[{{Vasiliev}(2019)}]{agama}
{Vasiliev}, E. 2019, \mnras, 482, 1525, \dodoi{10.1093/mnras/sty2672}

\bibitem[{{Vasiliev}(2023)}]{Vasiliev23}
---. 2023, \mnras, \dodoi{10.1093/mnras/stad2612}

\bibitem[{{Vasiliev} \& {Belokurov}(2020)}]{Vasiliev+20}
{Vasiliev}, E., \& {Belokurov}, V. 2020, \mnras, 497, 4162, \dodoi{10.1093/mnras/staa2114}

\bibitem[{{Vasiliev} {et~al.}(2021){Vasiliev}, {Belokurov}, \& {Erkal}}]{Vasiliev+21}
{Vasiliev}, E., {Belokurov}, V., \& {Erkal}, D. 2021, \mnras, 501, 2279, \dodoi{10.1093/mnras/staa3673}

\bibitem[{{Vislosky} {et~al.}(2024){Vislosky}, {Minchev}, {Khoperskov}, {Martig}, {Buck}, {Hilmi}, {Ratcliffe}, {Bland-Hawthorn}, {Quillen}, {Steinmetz}, \& {de Jong}}]{Vislosky+23}
{Vislosky}, E., {Minchev}, I., {Khoperskov}, S., {et~al.} 2024, \mnras, 528, 3576, \dodoi{10.1093/mnras/stae083}

\bibitem[{{Weinberg}(1992)}]{Weinberg92}
{Weinberg}, M.~D. 1992, \apj, 384, 81, \dodoi{10.1086/170853}

\bibitem[{{Weinberg} \& {Blitz}(2006)}]{Weinberg+Blitz06}
{Weinberg}, M.~D., \& {Blitz}, L. 2006, \apjl, 641, L33, \dodoi{10.1086/503607}

\bibitem[{{Widrow} {et~al.}(2012){Widrow}, {Gardner}, {Yanny}, {Dodelson}, \& {Chen}}]{Widrow+12_verticalwaves}
{Widrow}, L.~M., {Gardner}, S., {Yanny}, B., {Dodelson}, S., \& {Chen}, H.-Y. 2012, \apjl, 750, L41, \dodoi{10.1088/2041-8205/750/2/L41}

\bibitem[{{Williams} {et~al.}(2013){Williams}, {Steinmetz}, {Binney}, {Siebert}, {Enke}, {Famaey}, {Minchev}, {de Jong}, {Boeche}, {Freeman}, {Bienaym{\'e}}, {Bland-Hawthorn}, {Gibson}, {Gilmore}, {Grebel}, {Helmi}, {Kordopatis}, {Munari}, {Navarro}, {Parker}, {Reid}, {Seabroke}, {Sharma}, {Siviero}, {Watson}, {Wyse}, \& {Zwitter}}]{Williams+2013}
{Williams}, M.~E.~K., {Steinmetz}, M., {Binney}, J., {et~al.} 2013, \mnras, 436, 101, \dodoi{10.1093/mnras/stt1522}

\bibitem[{{Xu} {et~al.}(2015){Xu}, {Newberg}, {Carlin}, {Liu}, {Deng}, {Li}, {Sch{\"o}nrich}, \& {Yanny}}]{Xu+15_corrugations}
{Xu}, Y., {Newberg}, H.~J., {Carlin}, J.~L., {et~al.} 2015, \apj, 801, 105, \dodoi{10.1088/0004-637X/801/2/105}

\end{thebibliography}

\label{lastpage}
\end{document}